\documentclass[onecolumn, aps, pra, showpacs]{revtex4-1}
\pdfoutput=1

\usepackage{amssymb}
\usepackage{graphicx}
\usepackage{dcolumn}
\usepackage{bm}
\usepackage{amsmath}

\begin{document}

\title{Holographic interference in atomic photoionization from a
semiclassical standpoint}
\author{Sebasti\'{a}n D. L\'{o}pez}
\affiliation{Institute for Astronomy and Space Physics IAFE (UBA-Conicet), Buenos Aires,
Argentina}
\author{Diego G. Arb\'o}
\affiliation{Institute for Astronomy and Space Physics IAFE (UBA-Conicet), Buenos Aires,
Argentina}
\date{\today }

\begin{abstract}
A theoretical study of the interference pattern imprinted on the doubly
differential momentum distribution of the photoelectron due to atomic
ionization induced by a short laser pulse is developed from a semiclassical
standpoint. We use the semiclassical two-step model of Shvetsov-Shilovski
\textit{et al.} (Phys. Rev. A \textbf{94}, 013415) to elucidate the nature of the
holographic structure. Three different types of trajectories are
characterized during the ionization process by a single cycle pulse with
three different types of interferences. We show that the holographic
interference arises from the ionization yield only during the first half
cycle of the pulse, whereas the coherent superposition of electron
trajectories during the first half cycle and the second half cycle gives
rise to two other kinds of intracycle interference.
Although the picture of interference of a reference beam and a signal beam
is adequate, we show that our results for the formation of the holographic
pattern agree with the glory rescattering theory of Xia \textit{et al.} (Phys. Rev.
Lett. \textbf{121}, 143201). We probe the two-step semiclassical model 
by comparing it to the numerical results of the time dependent Schr\"{o}%
dinger equation.
\end{abstract}

\maketitle
\preprint{APS/123-QED}

%---------------------------------------------------

\section{\label{sec:level1}Introduction}

%---------------------------------------------------

The glory effect is a phenomenon found in many branches of physics. Firstly
observed in optics as a halo of one or more concentric rings around the
shadow of the observer, glories have been explained as the result of the
interference of the light entering droplets and following different paths 
\cite{Hulst47,Nussenzveig92,Berry15}. Many scattering processes in atomic
physics, like the decay of autoionizing states formed by the impact of slow
charged ions \cite{Swenson89,Swenson91,Barrachina89,Samengo96} and the anomalous
oscillations in the binary peak of electrons emitted in U$^{+21}$+He collisions,
have been explained as the interference of glory trajectories \cite{Reinhold91}.
Rainbow and glory scattering in Coulomb trajectories starting from a point
in space has been studied since the end of last century pointing out its importance
in atomic physics \cite{Swenson89,Samengo94,Samengo96,Cohen17,Kelvich16,Kelvich17}. 

Rescattering processes are responsible for different high energy structures
such as a plateau in the photoelectron energy spectrum \cite%
{Paulus94a,Paulus94b,Becker14,Luillier93,Paulus94,Paulus95,Becker02,Suarez15,Danek18a,Danek18b}
and the so-called rescattering rings in the momentum distributions \cite%
{Lewenstein95,Suarez15}. Although classical mechanics explains many features
of electron distributions in atomic photoionization \cite{Liu14}, electron
dynamics can only be fully described by quantum mechanics as quantum
interference effects. Spatial and temporal interferences have been studied
both experimentally and theoretically. Gribakin and Kuchiev first reported
quantum interference within an optical cycle in Ref. \cite{Gribakin97} and
Paulus \textit{et al.} observed and analyzed them theoretically for negative
ions. Chirila \textit{et al.} calculated non-equidistant peaks in the
photoelectron spectrum \cite{Chirila05}. A time double-slit interference
pattern has been measured \cite{Lindner05,Gopal09} and theoretically
studied \cite{Becker02,Arbo06b,Arbo10a,Arbo10b,Arbo12,Yang16} for few-cycle
pulses. A bouquet shape structure in the doubly differential momentum 
distribution near threshold was measured and understood as the
interference of electron trajectories oscillating around Kepler hyperbolae
in a generalized Ramsauer-Townsend scheme \cite%
{Rudenko04,Maharjan06,Arbo06a,Arbo08b,Borbely13,Yan10}.

In the last decade, some structures coming from interference of rescattered
electrons with those which ionize without returning to the parent ion were
characterized as holographic structures in photoelectron spectra \cite%
{Huismans11,Huismans12,Song16,Lai17,Shilovski18,Tan19}. Electron holography
is useful for probing some properties of the ionization process. In this
sense, Porat \textit{et al.} performed an experiment showing the detailed
sub-cycle electron dynamics associated with the hologram \cite{Porat18}.
Very recently, Xia \textit{et al.} explained the holographic structure found
in the electron momentum distribution in the strong-field atomic ionization
as the result of quantum interference of glory rescattering semiclassical
trajectories \cite{Raz12,Xia18}. As a spider-like shape in the doubly
differential momentum distribution, holographic interference is one of many
types of interferences visible in experiments of atomic and
molecular ionization by laser pulses with frequencies in the far infrared 
\cite{Huismans11,Huismans12,Xie16}. However, for ionization by infrared and
near infrared (NIR) lasers, the holographic interference pattern can 
hardly been seen in the electron yield.

In this work, we explore the nature of subcycle dynamics of the atomic
ionization by a NIR single-cycle laser pulse leading to the holographic
pattern in the momentum distribution within a semiclassical theory by using
the semiclassical two-step model (SCTS \cite{Shilovski16,Song17,Ni18}) and
compare the results with a pure quantum treatment. We show that glory trajectories
present in the forward direction are in the transition region between
rescattering and direct trajectories. Besides the holographic pattern, we
show two other types of intracycle interference also present in the doubly
differential photoelectron momentum distribution: The well-known intracycle
interference, stemming from the interference of non-rescattering
(direct and indirect) electron trajectories \cite%
{Chirila05,Arbo06a,Arbo06b,Xiao16}, and the intracycle interference,
stemming from direct and rescattering trajectories \cite%
{Zagoya14,Maxwell18a,Maxwell18b}.

The paper is organized as follows. In Sec. \ref{theory} we present the
semiclassical two-step (SCTS) model used to analyze the different
interfering types of electron trajectories and, thus, different kinds of
interferences present in the photoionization process. We also mention our
method to numerically solve the time dependent Schr\"{o}dinger equation
(TDSE) \cite{Tong97,Tong00,Tong05} and briefly pose the glory rescattering
theory (GRT) \cite{Xia18}. In Sec. \ref{results} we show and discuss our
results of the different interference structures, especially the holographic
structure in view of an interference process of glory rescattering trajectories.
Finally, in Sec. \ref{conclusions} we draw the fundamental concluding remarks.

We employ atomic units throughout this work.

%---------------------------------------------------

\section{\label{theory}Theory}

In the length gauge, the Hamiltonian of an atomic system interacting with a
laser pulse within the single-active electron approximation can be written as%
\begin{equation}
H=\frac{\vec{p}{\,}^{2}}{2}+V(r)+\vec{r}\cdot \vec{F}\,(t),  \label{hami}
\end{equation}%
where the first term corresponds to the kinetic energy of the active
electron with electron momentum $\vec{p}$, the electron position from the
atomic core is $\vec{r},$ and $V(r)$ is the time-independent central
potential of the core composed by the atomic nucleus and the rest of the
electrons considered frozen. The summation of these two terms forms the
time-independent Hamiltonian of the atom. The last term in the right-hand
side of Eq. (\ref{hami}), $\vec{r}\cdot \vec{F}\,(t)$, describes the
interaction of the atomic system with the time-dependent electric field $%
\vec{F}\,(t)$ of the laser pulse within the dipole approximation.

The photoelectron momentum distribution after photoionization can be
calculated as

\begin{equation}
\frac{dP}{d\vec{k}}=\left\vert T\right\vert ^{2},
\end{equation}
where $T$ is the transition matrix from the initial bound state to the final
state of an electron with momentum $\vec{k}$ in the continuum. There are
many ways to calculate the transition matrix from pure classical to quantum
calculations with several levels of approximation. In this paper, we focus
on the study of the semiclassical two-step model (SCTS) firstly introduced
in \cite{Shilovski16} based on classical trajectory Monte Carlo models that
include quantum interferences \cite{Shilovski12,Li14} and compare the
results with the \textit{ab initio} solution of the time dependent Schr\"{o}%
dinger equation (TDSE) \cite{Tong97,Tong00,Tong05}. In the rest of the
section we briefly describe both calculating methods together with the glory
rescattering theory of Xia \textit{et al.} \cite{Xia18}.

\subsection{Semiclassical model}

%---------------------------------------------------

Here we briefly describe the SCTS. For a thorough description of the model
and its theoretical framework, the reader can refer to Ref. \cite%
{Shilovski16}. The method assumes that the ionization process of the atom
happens in two different steps. The first step is the tunneling through the
potential barrier formed by the atomic central potential $V(r)$ and the
interaction energy with the external field, $\vec{r}\cdot \vec{F}\,(t)$,
corresponding to the last two terms of the Hamiltonian in Eq. (\ref{hami}).
The second step corresponds to the action of the Coulomb force $-\partial
V(r)/\partial r$ and the external field $\vec{F}(t)$ on the electron in the
continuum.

The time-dependent distorted wave theory establishes that the transition
amplitude in the prior form and length gauge is expressed as \cite%
{Dewangan94,Macri03}

\begin{equation}
T=-i\int_{-\infty }^{+\infty }dt\left\langle \chi _{\vec{k}}^{-}(\vec{r}%
,t)\right\vert \vec{r}\cdot \vec{F}(t)\left\vert \phi _{i}(\vec{r}%
,t)\right\rangle \,  \label{Tif}
\end{equation}%
where $\phi _{i}(\vec{r},t)=\varphi _{i}(\vec{r})\,\mathrm{e}^{\mathrm{i}%
I_{p}t}$ is the initial atomic state with ionization potential $I_{p}$ and $%
\chi _{\vec{k}}^{-}(\vec{r},t)$ is the distorted final state.

The time integral in Eq. (\ref{Tif}) can be calculated with the saddle-point
approximation if the phase of the integrand, i.e., the action $\Phi (\vec{k}%
,t)=\mathrm{Arg}\left[ \left\langle \chi _{\vec{k}}^{-}(\vec{r}%
,t)\right\vert \vec{r}\cdot \vec{F}(t)\left\vert \phi _{i}(\vec{r}%
,t)\right\rangle \right] $ varies rapidly with time. This is the so-called
semiclassical approximation, which states that the action in the Feynman
propagator is asymptotically large compared to the quantum action $\hbar$
and consequently assures the use of the saddle point approximation.
In this way, the transition matrix becomes a sum over
several electron trajectories born at ionization times $t_{s},$ i.e.,%
\begin{equation}
T=\sum_{t_{s}}\vec{F}(t_{s})\int d\vec{r}\frac{e^{i\Phi \left( \vec{k}%
,t_{s}\right) }}{\ddot{\Phi}\left( \vec{k},t_{s}\right) }|\chi _{\vec{k}%
}^{-\ast }(\vec{r},t_{s})|\vec{r}\varphi _{i}(\vec{r}),  \label{T-saddle}
\end{equation}%
with the dipole element $\vec{d}(\vec{k},t_{s})=\left\vert \left\langle \chi
_{\vec{k}}^{-}(\vec{r},t_{s})\right\vert \vec{r}\left\vert \phi _{i}(\vec{r}%
,t_{s})\right\rangle \right\vert $ containing the spatial dependencies of
the transition amplitude. The saddle points $t_{s}$ correspond to the
ionization times in the complex plane and fulfill the saddle equation $\dot{%
\Phi}\left( \vec{k},t_{s}\right) =0,$ where the dot and double dot on the
action mean that the respective time derivative and double time derivative
must be taken.

The imaginary part of $t_{s}$ produces an exponential decay in the
probability corresponding to the first step in our semiclassical
description. For the first step, the strong field approximation
(SFA), which neglects the Coulomb distortion in the final channel, is considered.
With this in mind, the final distorted function is a Volkov state and the
dipole element becomes $\vec{d}(\vec{k},t_{s})=\left\langle \vec{k}+\vec{A}%
(t_{s})\right\vert \vec{r}\left\vert \varphi _{i}(\vec{r})\right\rangle ,$
where the bra corresponds to a plane wave. Therefore, the action becomes the
generalized Volkov action which includes the energy of the initial state $%
-I_{p}$ \cite{Volkov35}
\begin{equation}
\Phi \left( \vec{k},t_{s}\right) =\left[ \vec{k}+\vec{A}(t)\right] \cdot 
\vec{r}-\int_{t}^{\infty }dt^{\prime }\left[ \frac{\left( \vec{k}+\vec{A}%
(t^{\prime })\right) ^{2}}{2}+I_{p}\right] .  \label{Volkov-phase}
\end{equation}
This leads to the very well-known PPT (Perelomov-Popov-Terent'ev) or ADK
(Ammosov-Delone-Krainov) tunneling rates \cite{Landau65,PPT66,ADK86}
\begin{equation}
w_{0}\left( t_{0},v_{0\perp }\right) \propto e^{-\frac{2\left( 2I_{P}\right)
^{3/2}}{3F\left( t_{0}\right) }}e^{-\frac{\sqrt{2I_{P}}v_{0\perp }^{2}}{%
F\left( t_{0}\right) }},  \label{tunnel-rate}
\end{equation}
where $t_{0}=\rm{Re} \lbrack t_{s}\rbrack$, and $v_{0\perp }$ refers to the velocity
in the direction perpendicular to the polarization axis at time $t_o$.
The electron is supposed to tunnel through the barrier formed by 
$V(r)+\vec{r}\cdot \vec{F}\,(t)$ instantaneously 
(in the complex plane from complex times $t_{s}$ to real times $t_{0}$)
with zero longitudinal probability $v_{0,z}$ and a
Gaussian distributed probability $v_{0\perp },$ according to Eq. (\ref%
{tunnel-rate}). The assumption $v_{0,z}=0$ is not strictly fulfilled for $%
t_{0}$ different from extremes of the electric field $F(t)$, which leads to
non-adiabatic effects that we neglect in this paper. For the
coordinates right after tunneling we use $z_{0}=-\sqrt{I_{P}/F\left(
t\right) }$ (the semiclassical distance traveled under the barrier for a
zero range potential) and zero perpendicular coordinate. In our simulations
we neglect the Stark shift of the initial state. The initial conditions for
the second step are the position and momentum distributions in the phase
space right after the first step. We use an acceptance-rejection algorithm in
order to reproduce the initial distribution.

The second step consists in simulating the time evolution of the system
classically by solving the Hamilton's equations of motion%
\begin{equation}
\overset{\cdot }{\vec{r}}=\frac{\partial H}{\partial \vec{p}};\qquad -%
\overset{\cdot }{\vec{p}}=\frac{\partial H}{\partial \vec{r}},
\label{hamilton-eqs}
\end{equation}%
where the Hamiltonian $H$ is given by Eq. (\ref{hami}). The first of
the Eqs. (\ref{hamilton-eqs}) expresses that the momentum is equal to the
velocity (in atomic units), i.e., $\overset{\cdot }{\vec{r}}=\vec{p}$ in the
length gauge, whereas the second one leads to the second Newton's law
$\overset{\cdot }{\vec{p}}=-\partial V(r)/\partial \vec{r}-\vec{F}(t)$.
The SFA neglects the potential energy between the remaining core and the
active electron (the first term of the second-hand side of Newton's law),
however, we keep it in the time evolution of each electron trajectory
during the second step of the photoionization process. The electron evolves
under the Hamilton's equations [Eqs. (\ref{hamilton-eqs})] acquiring a phase
given by the classical action along the evolution from $t_0$ up to the detection
time. Then, the probability amplitude is accounted as the coherent superposition 
of the phases $\Phi $ of each electron trajectory according to Eq.
(\ref{T-saddle}) replacing the saddle times by the ionization times $t_{0}$.

For calculating the phases we need to consider the matrix element of the
semiclassical propagator between the initial state at time $t_{0}^{j}$ (for
the $j$th trajectory) and the final state at time $t\rightarrow \infty $
(the time that the electron impinges on the detector, which
compared to the atomic transition times can be regarded as infinite).
The photoionization is a half-scattering process of an electron
initially located in the vicinity of the ionic core at real time $t_{0}$
and measured with final momentum $\vec{k}$ at the detector ($t\rightarrow \infty $).
Therefore, the classical phase is associated with the integral of the Lagrangian
through a Legendre transformation \cite{Goldstein02,Miller71,Walser03,Spanner03}, i.e,
\begin{equation}
\Phi \left( \vec{k},t_{0}^{j}\right) =\left[ \vec{k}+\vec{A}(t)\right] \cdot 
\vec{r}+\int_{t_{0}^{j}}^{\infty }dt\left[ \vec{p}(t)\cdot \overset{\cdot }{%
\vec{r}}(t)-H\right] +I_{p}t_{0}^{j}-\vec{k}\cdot \vec{r}(t\rightarrow
\infty ).  \label{phi3}
\end{equation}
Integrating the second term in the right hand side of Eq. (\ref{phi3})) by parts
and performing some approximations from Feynman propagators
(see \cite{Shilovski16} for a complete discussion),
the phase can be expressed as  
\begin{equation}
\Phi \left( t_{0}^{j},\mathbf{v}_{0}^{j}\right) =\left[ \vec{k}+\vec{A}(t)%
\right] \cdot \vec{r}+I_{P}t_{0}^{j}-\vec{v}_{0}^{j}\cdot \vec{r}%
_{0}^{j}-\int_{t_{0}^{j}}^{\infty }dt\left( \frac{\vec{p}^{2}(t)}{2}+V(r)-%
\vec{r}(t)\cdot \frac{\partial V(r)}{\partial \vec{r}}\right) ,
\label{phase-atom}
\end{equation}%
where $\vec{r}_{0}^{j}$ is the initial position (at time $t_{0}^{j}$) of the 
$j$th trajectory resulting from the first step. The last term in the
integrand of Eq. (\ref{phase-atom}) is completely neglected in the
quantum trajectory Monte Carlo (QTMC) model \cite{Li14}.
For a hydrogenic case, i.e., $V(r)=-Z/r,$ the phase in Eq. (\ref{phase-atom})
becomes
\begin{equation}
\Phi \left( t_{0}^{j},\mathbf{v}_{0}^{j}\right) =\left[ \vec{k}+\vec{A}(t)%
\right] \cdot \vec{r}+I_{P}t_{0}^{j}-\vec{v}_{0}^{j}\cdot \vec{r}%
_{0}^{j}-\int_{t_{0}^{j}}^{\infty }dt\left( \frac{\vec{p}^{2}(t)}{2}-\lambda 
\frac{Z}{r(t)}\right) ,  \label{phase-h}
\end{equation}
with $\lambda =2.$ We refer to Eq. (\ref{phase-h}) with $\lambda =2$ to the
SCTS phase. In our simulations the third term in Eq. (\ref{phase-h}) is zero
since $\vec{r}_{0}^{j}=$ $-\sqrt{I_{P}/F\left( t\right) }\hat{z}$ and we
consider that the velocity right after tunneling is perpendicular to the
polarization direction of the laser field. In turn, the QTMC model considers
the phase as in Eq. (\ref{phase-h}) with $\lambda =1,$ which is a first
order approximation of the SCTS phase \cite{Li14}.

In order to numerically implement the second step, we divide the time
evolution into two different intervals: From the initial
time of the $j$th trajectory to the end of the laser pulse of duration $\tau
$, i.e., $[t_{0}^{j},\tau ]$, and from the end of the pulse to the asymptotic time
$t\rightarrow\infty$, i.e., $[\tau ,\infty )$.
It is worth noting that for a hydrogenic atom,
during the second time interval when the external laser field is off,
the different electron trajectories follow Kepler trajectories up to the
detector and the contribution to the phase can be taken into account
analytically without performing the numerical evolution of the electron \cite%
{Shilovski12,Shilovski16}. Therefore, the asymptotic momentum can be
calculated as%
\begin{equation}
\vec{k}=\frac{k^{2}\left( \vec{L}\times \vec{a}\right) -kZ\vec{a}}{%
Z^{2}+k^{2}L^{2}},  \label{asym-mom}
\end{equation}%
where the absolute value of the asymptotic momentum $k$ is related to the
absolute value of the momentum at time $t=\tau$ through the conservation
of the energy, i.e., $k^{2}/2=p^{2}(\tau )/2-Z_{T}/r(\tau )$.
The Runge-Lenz vector can be determined as 
$\vec{a}=\vec{p}(\tau )\times \vec{L}-Z\ \vec{r}(\tau)/r(\tau ),$ and $\vec{L}$
is the angular momentum (which is also a constant of motion)
after the laser has been switched off.

As the time extends to infinity, the integral in the phases in Eq. (\ref%
{phase-h}) contains divergent terms. For that reason, the integral is split
at the instant corresponding to the end of the pulse $\tau $ as
\begin{equation}
\Phi \left( t_{0}^{j},\mathbf{v}_{0}^{j}\right) =\left[ \vec{k}+\vec{A}(t)%
\right] \cdot \vec{r}+I_{P}t_{0}^{j}-\vec{v}_{0}^{j}\cdot \vec{r}%
_{0}^{j}-\int_{t_{0}^{j}}^{\tau }\left( \frac{p^{2}}{2}-\lambda \frac{Z}{r(t)}%
\right) dt+\left( \lambda -1\right) \phi _{C},  \label{phase}
\end{equation}
where
\begin{equation}
\phi _{C}=-\int_{\tau }^{\infty }\frac{Z}{r(t)}dt.  \label{coulomb-phase}
\end{equation}%
In Eq. (\ref{phase}), we have dropped the diverging energy term $\int_{\tau
}^{\infty }\left[ p^{2}/2-Z/r(t)\right] dt=$ $\int_{\tau }^{\infty }k^{2}/2dt
$ because it is the same for all trajectories with the same final
momentum. In contrast to the SCTS ($\lambda =2$), the QTMC model lacks the
asymptotic Coulomb correction to the phase given by the last term in Eq. (%
\ref{phase}) since $\lambda =1$ and, thus, $\Phi ^{QTMC}$ remains exactly as
was stated in Ref. \cite{Li14}.

The asymptotic Coulomb phase $\Phi_c$ in Eq. (\ref{coulomb-phase}) is still divergent.
It can be regularized by a change of coordinates $r(t)=(e\cosh \xi -1)/(2E)$,
where $e=\sqrt{k^{2}L^{2}+Z^{2}}$ is the eccentricity of the Kepler orbit and
$\xi=\xi (t)$ is determined from $t=(e\sinh \xi -\xi )/(2E)^{3/2}+C,$ where $C$
can be found from the position and velocity at $t=\tau .$ With this in mind,
Eq. (\ref{coulomb-phase}) becomes
\begin{equation}
\phi _{C}=\frac{Z}{\sqrt{2E}}\left[ \xi (\infty )-\xi (\tau )\right] ,
\label{coulomb-phase2}
\end{equation}
where $\xi (\infty )$ means that the limit $t\rightarrow \infty $ of $\xi (t)
$ should be taken. In fact, this is the divergent part of the Coulomb phase
in Eq. (\ref{coulomb-phase}). In this sense, we can neglect the constant $C$
and also $\xi $ compared to $\sinh \xi $ and the time can be asymptotically
written as $t=e\exp (\xi )/(2E)^{3/2}$, or equivalently $\xi (t)=\ln \left[
(2E)^{3/2}t/e\right] =\ln \left[ (2E)^{3/2}t\right] -\ln \left[ e\right] .$
For all the trajectories with the same final momentum $\vec{k},$ the first
term of $\xi (t)$ is the same, thus we drop it off in our calculations. In
turn, the second term depends on the energy and angular momentum through the
eccentricity parameter $e$ and, contrarily to the energy, the angular
momentum is in general different for all the interfering trajectories with
the same final momentum $\vec{k}.$ From the expression of $r(t),$ we can
write $\xi (\tau )=\pm \cosh^{-1}\left[ \frac{2E\ r(\tau )+1}{e}\right]$.
With a bit of algebra, the second term in the Eq. (\ref{coulomb-phase2})
can be written as 
$\xi (\tau )=\sinh^{-1}\left[\sqrt{2E}\frac{\vec{r}(\tau )\cdot \vec{p}(\tau )}{e}\right]$
and, therefore, the interference contribution to the Coulomb phase reads
\begin{equation}
\phi _{C}=-\frac{Z}{\sqrt{2E}}\left\{ \ln e+\sinh^{-1}\left[ \frac{%
\sqrt{2E}}{e}\vec{r}(\tau )\cdot \vec{p}(\tau )\right] \right\} .
\label{coulomb-phase3}
\end{equation}

Now that the Coulomb correction of the phase (and thus, the phase itself)
has been properly accounted, Eq. (\ref{T-saddle}) is computed together with
the SFA assumption for the first step in Eq. (\ref{tunnel-rate}). The
ionization probability can then be calculated as
\begin{equation}
\frac{dP}{d\vec{k}}=|T|^{2}=\left\vert \sum_{j}\sqrt{w_{0}\left(
t_{0}^{j},v_{0\perp }^{j}\right) }e^{i\Phi \left( \vec{k},t_{0}^{j}\right)
}\right\vert ^{2},  \label{probability}
\end{equation}
where the sum extends over all electron trajectories. The CTMC approximation
is reached when all the phases are neglected by randomizing their values.
Therefore, the CTMC ionization probability becomes
\begin{eqnarray}
\frac{dP}{d\vec{k}} &=&\left\vert \sum_{j}\sqrt{w_{0}\left(
t_{0}^{j},v_{0\perp }^{j}\right) }e^{i\Phi \left( \vec{k},t_{0}^{j}\right)
}\right\vert ^{2}  \notag \\
&=&\sum_{j}\sum_{j^{\prime }}\sqrt{w_{0}\left( t_{0}^{j},v_{0\perp
}^{j}\right) }\sqrt{w_{0}\left( t_{0}^{j^{\prime }},v_{0\perp }^{j^{\prime
}}\right) }e^{i\left[ \Phi \left( \vec{k},t_{0}^{j}\right) -\Phi \left( \vec{%
k},t_{0}^{j^{\prime }}\right) \right] }  \notag \\
&=&\sum_{j}w_{0}\left( t_{0}^{j},v_{0\perp }^{j}\right) ,  \label{CTMC-proba}
\end{eqnarray}
where the exponential on the second line of Eq. (\ref{CTMC-proba}) takes all
random values if $j\neq j^{\prime }$ and zero if $j=j^{\prime }.$ Therefore,
all crossed terms in the second line go to zero as the number of trajectories
goes to infinity because of the randomness of the phase. Therefore, only the
terms with $j=j^{\prime }$ survive and the final CTMC probability
distribution is finally found in the third line of Eq. (\ref{CTMC-proba}).

In our calculations, we use importance sampling to compute Eq. (\ref%
{probability}), where the weight $\sqrt{w_{0}\left( t_{0}^{j},v_{0\perp
}^{j}\right) }$ of a given trajectory is already considered at the sampling
stage by choosing the initial sets of initial conditions $t_{0}^{j}$ and $%
\vec{v}_{0}^{j}$ distributed taking into account the tunneling probability
in Eq. (\ref{tunnel-rate}). In this way, the electron distribution can be
written simply as
\begin{equation}
\frac{dP}{d\vec{k}}=\left\vert \sum_{j}e^{i\Phi \left( \vec{k}%
,t_{0}^{j}\right) }\right\vert ^{2},
\end{equation}
and, consequently, less number of trajectories is needed to reproduce the
interference structures compared to using uniformly
distributed initial conditions.

\subsection{Glory rescattering theory}

As the semiclassical model states, the semiclassical transition amplitude is given by 
the sum over all classical trajectories starting at exit position $\vec{r}(t_{0})$
Eq. (\ref{phi3}) \cite{Miller71,Kay05} and can be
written as [Eq. (\ref{T-saddle})]
\begin{equation}
T=\sum_{t_{0}}\int d\vec{r}\sqrt{w_{0}\left( t_{0},v_{0\perp }\right) }\frac{%
e^{i\Phi \left( \vec{k},t_{0}\right) }}{\ddot{\Phi}\left( \vec{k}%
,t_{0}\right) }.  \label{T-saddle2}
\end{equation}

The spatial integration in Eq. (\ref{T-saddle2}) is generally solved using
the saddle point approximation. In turn, following the derivation
in the supplemental material of Ref. \cite{Xia18}, as the photoionizing
system possesses cylindrical symmetry around the polarization axis,
the integral in Eq. (\ref{T-saddle2}) can be solved in cylindrical 
coordinates as
\begin{equation}
T=\sum_{t_{0}}\int d\rho \ \rho \int dz\frac{\sqrt{w_{0}\left(
t_{0},v_{0\perp }\right) }}{\ddot{\Phi}\left( \vec{k},t_{0}\right) }\int
d\phi e^{i\Phi \left( \vec{k},t_{0}\right) }.
\end{equation}
It is invalid to apply the steepest descend method over the azimuth angle $%
\phi $ because of the presence of an axial singularity \cite{Xia18}. The
angular integral can be performed analytically as%
\begin{eqnarray}
\int d\phi e^{i\Phi \left( \vec{k},t_{0}\right) } &\varpropto &\int d\phi
e^{i\vec{p}\cdot \vec{r}}=e^{ip_{z}\cdot z}\int d\phi e^{ip_{x}\rho \cos
\phi +ip_{y}\rho \sin \phi }  \label{azimuthat-integral} \\
&\varpropto &e^{ip_{z}\cdot z}\int d\phi e^{ip_{x}\rho \cos \phi
}=e^{ip_{z}\cdot z}J_{0}(p_{x}\rho )  \notag
\end{eqnarray}%
without losing generality, in the right hand side of Eq. (\ref%
{azimuthat-integral}) we set $p_{y}=0$ due to cylindrical symmetry which
leads to the Bessel function of the first kind in the second line. Thus, for
the axial singularity and using the saddle point approximation for the
radial and longitudinal coordinates $\rho $ and $z,$ we finally find that 
\cite{Xia18}%
\begin{equation}
T\sim \sqrt{w_{0}\left( t_{0},v_{0\perp }\right) p_{\rho 0}b}J_{0}(k_{\perp
}b),  \label{GRT}
\end{equation}%
where $b$ and $v_{\perp 0}$ are the asymptotic impact parameter and the
initial transverse momentum.

\subsection{Time dependent Schr\"odinger equation}

In order to numerically solve the TDSE in the dipole approximation with the
Hamiltonian given by Eq. (\ref{hami}), we employ the generalized
pseudo-spectral method, which combines the discretization of the radial
coordinate optimized for the Coulomb singularity with quadrature methods to
allow stable long-time evolution using a split-operator representation of
the time-evolution operator \cite{Tong97,Tong00,Tong05}. Both the bound as
well as the unbound parts of the wave function $|\psi _{\vec{k}}(t)\rangle $
can be accurately represented. Due to the cylindrical symmetry of the system
the magnetic quantum number $m$ is conserved. After the end of the laser
pulse the wave function is projected on eigenstates $|k,\ell \rangle $ of
the free atomic Hamiltonian with positive eigenenergy $E=k^{2}/2$ and
orbital quantum number $\ell $ to determine the transition amplitude $T$
to reach the final state $|\phi _{f}\rangle $ (see Refs. \cite%
{Schoeller86,Messiah73,Dionissopoulou97}). In order to avoid unphysical
reflections of the wave function at the boundary of the system, the length
of the computing box was chosen to be 1200 a.u.\ ($\sim 65$ nm) and the
maximum angular momentum considered was $\ell _{\max }=200$.

\section{Results and discussion}

\label{results} %--------------------------------------------------------

For the sake of simplicity, throughout the paper we use a linearly polarized 
single-cycle laser pulse
\begin{equation}
\vec{F}(t)=F_{0}\sin \omega t\ \hat{z}  \label{field}
\end{equation}
for $0\leq t\leq 2\pi /\omega $ and zero elsewhere. We
use a peak field $F_{0}=0.075$ a.u., which corresponds to a laser intensity
of $I=2\times 10^{14}$ W/cm$^{2}$, and a laser frequency $\omega =0.05$ a.u.,
corresponding to a wavelength of $911$ nm, very close to the Ti-Saphire
laser frequency. As the system possesses cylindrical symmetry around the
polarization axis $\hat{z},$ the ionization process can be thought as a
two-dimensional problem where the projection of the angular momentum of the
electron along the polarization axis is conserved,
i.e., the magnetic quantum number is constant.

%--------------------------------------------------------------------------
\begin{figure}[tbp]
\centering
\includegraphics[width=0.8\textwidth]{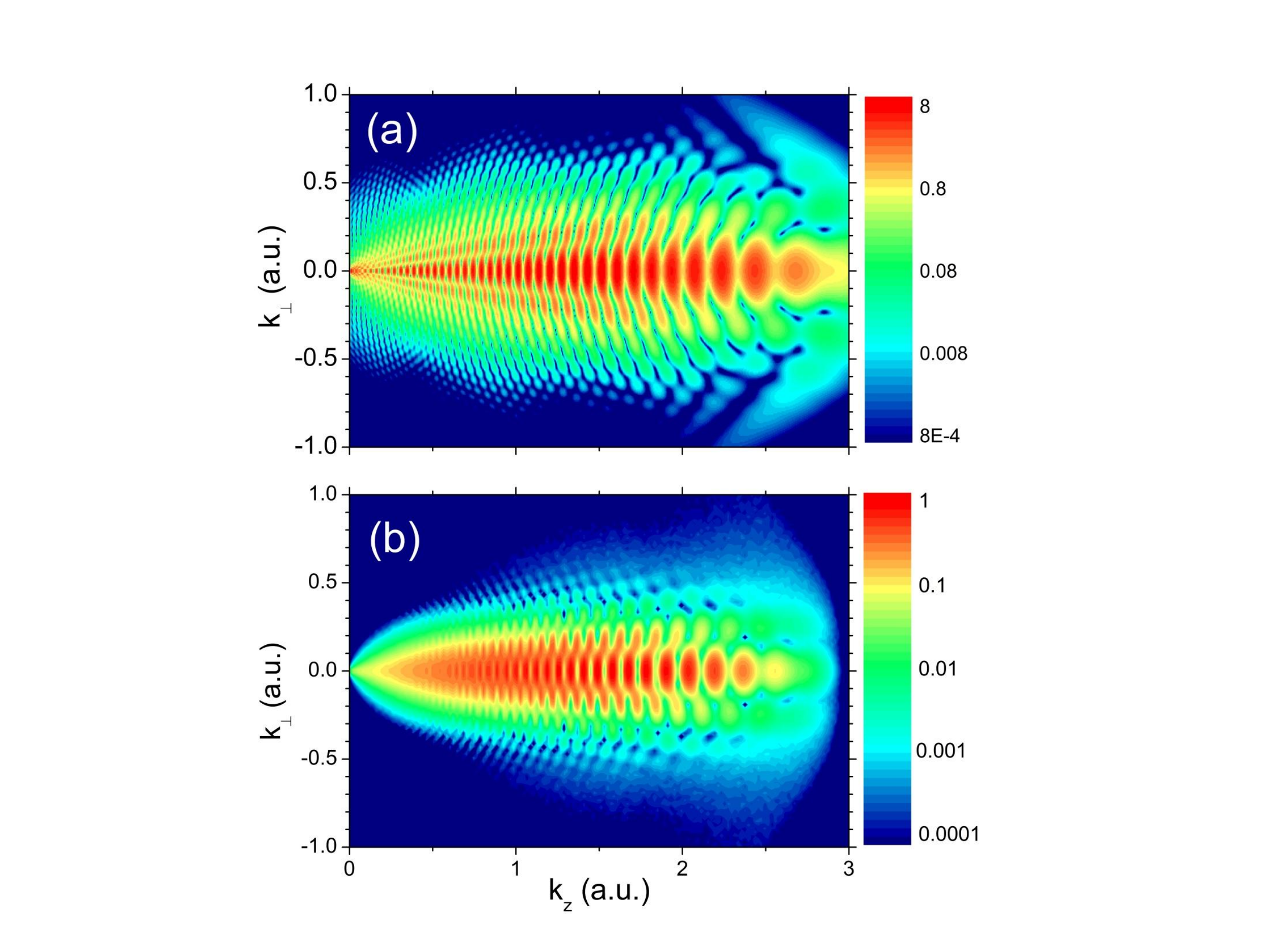}
\caption{Doubly differential momentum distribution for ionization of
atomic hydrogen by the one-cycle sine pulse of Eq. (\ref{field}) with
$F_{0}=0.075,$ $\omega =0.05$ calculated within the (a) TDSE and (b) SCTS.
}
\label{full-mom-dist}
\end{figure}
%--------------------------------------------------------------------------

For the single-cycle electric field of Eq. (\ref{field}), the simple man's
model (SMM) predicts ionization only in the forward direction, i.e, $%
0<k_{z}<2F_{0}/\omega $ (see, for example \cite{Arbo06b}). If one wants to
obtain forward-backward symmetrical ionization like in experiments, one
needs to use longer electric fields with some ramp on and ramp off.
However, we show below that using the single-cycle pulse of Eq. (\ref{field}%
) is sufficient to show most of the interference processes characteristic of
the electron yield for a more realistic laser pulse. The electron yield
after ionization of atomic hydrogen by the electric field in Eq. (\ref{field}%
) calculated within the TDSE can be seen in Fig. \ref{full-mom-dist}a as a
function of the longitudinal momentum $k_{z}$ (along the polarization
direction) and transverse momentum $k_{\perp }$ (perpendicular to the
polarization direction). The momentum distribution in Fig. \ref%
{full-mom-dist}a spreads mostly along the forward direction and within the
classical boundaries predicted by the SMM, i.e., $0<k_{z}<3,$ though
extending slightly beyond the classical boundaries due to quantum diffusion.
We can see a very rich interference pattern in the quantum momentum
distribution. In order to perform the identification of the different kinds
of interference present in the complicated interference pattern of Fig. \ref%
{full-mom-dist}a, we also compute the ionization of the hydrogen atom within
the SCTS model of Sec. \ref{theory} A using the same electric field of Eq. (%
\ref{field}). The SCTS simulation was performed in the four-dimensional
phase space $(z,x,v_{0z},v_{x0})$, where $x$ is the component of the
position perpendicular to the laser direction. In cylindrical coordinates,
we should note that $\rho =|x|.$ We observe that the SCTS distribution in
Fig. \ref{full-mom-dist}b restricts to the classical boundaries, as
expected. We can see that both classical and quantum momentum distributions
exhibit a similar interference pattern, although the resemblance is not
perfect.

%--------------------------------------------------------------------------
\begin{figure}[tbp]
\centering
\includegraphics[width=0.6\textwidth]{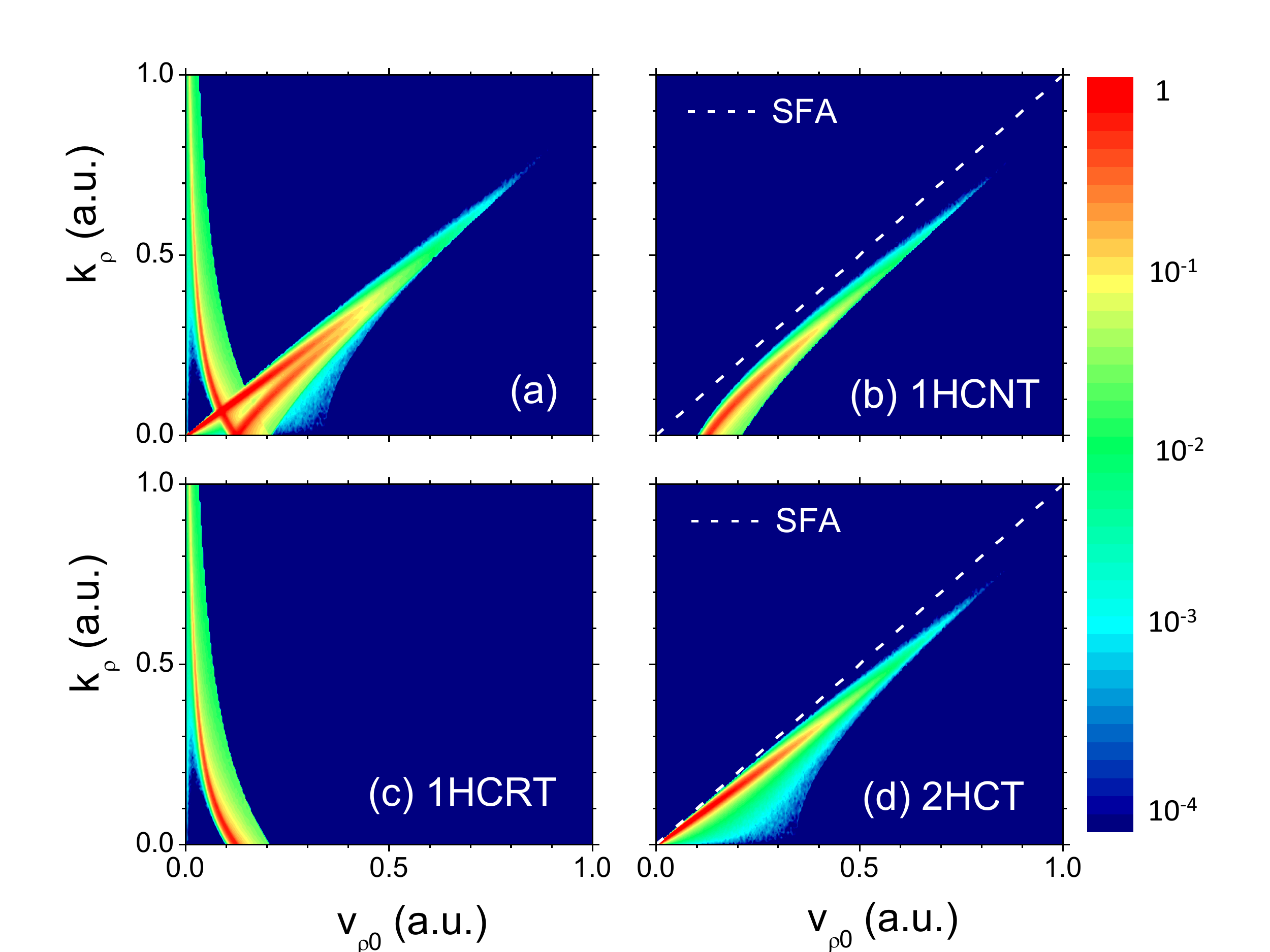}
\caption{Maps of final perpendicular momentum versus initial
perpendicular momentum (right after ionization) for (a) all trajectories,
(b) 1HCNT (non-rescattering trajectories ionized during the first half cycle),
(c) 1HCRT (rescattering trajectories ionized during the first half cycle),
and (d) 2HCT (trajectories ionized during the second half cycle).
The dashed line corresponds to the SFA $k_{\rho }=v_{\rho 0}$ prediction.
The laser parameters are the same as in Fig. \ref{full-mom-dist}.
}
\label{krho-vrho0}
\end{figure}
%--------------------------------------------------------------------------

In order to analyze the different types of electron trajectories present in
the ionization process, we show in Fig. \ref{krho-vrho0}a the map of
asymptotic (final) perpendicular momenta (in cylindrical coordinates) $%
k_{\rho }=|k_{\perp }|$ versus the momenta at the time of ionization (also in
cylindrical coordinates) $v_{0\rho }=|v_{0\perp }|$ calculated for a total
of about 20 million trajectories within the CTMC [see Eq. (\ref{CTMC-proba})].
Within the SFA, the perpendicular momentum is constant because the action
of the ionic potential of the remaining core on the escaping electron is neglected,
i.e., $k_{\rho}=v_{\rho 0},$ which is indicated as a dashed line in Figs. 
\ref{krho-vrho0}b and d. Beyond the SFA, electron trajectories can be classified
according to the effect of the Coulomb potential on them: (i) weak effect,
where the Coulomb potential is not strong enough to change the perpendicular
direction of the electron trajectories, i.e., $v_{\perp 0}k_{\perp }>0$,
that we call non-rescattering trajectories, also known in the literature as
farside trajectories \cite{Fuller75,Samengo96}, and (ii) strong effect,
where the Coulomb potential is strong enough to change the perpendicular
direction of the electron trajectories, i.e., $v_{\perp 0}k_{\perp }<0$,
that we call rescattering trajectories, also known in the literature as
nearside trajectories \cite{Fuller75,Samengo96}. We can see three different regions 
in the $(k_{\rho },v_{\rho 0})$ map of Fig. \ref{krho-vrho0}a. The 
$(k_{\rho},v_{\rho 0})$ map in Fig. \ref{krho-vrho0}b shows only the non-rescattering
trajectories ionized during the first half cycle, which we call 1HCNT. One
can see very clearly the weak effect of the ionic potential on the escaping
trajectories slowing down the electron in the perpendicular direction, i.e., 
$k_{\perp }<v_{\perp 0}.$ In our case of the hydrogen atom, this effect is
called Coulomb focusing, although this name is commonly
extended to other atoms with non-Coulombic potentials \cite%
{Kelvich16,Danek18a,Danek18b}. We see that 1HCNT must have an initial
transversal momentum higher than a value between $0.1$ a.u. and $0.2$ a.u.
for our case. Electron trajectories with less than these values for the
initial perpendicular velocity ionized during the first half cycle are
strongly affected by the potential of the remaining core changing the sign
of the transverse momentum, which can be seen as a collision of the
escaping electron with the parent ion. The $(k_{\rho },v_{\rho 0})$ map for
this kind of trajectories, that we name 1HCRT, is shown in Fig. \ref%
{krho-vrho0}c. Perpendicular initial velocities are low enough so that the
action of the Coulomb potential is strong enough to change the sign of $%
v_{0x}$ and produce rescattering. We observe that for electron trajectories with
low initial transversal momentum (less than $0.1-0.2$ a.u.), the asymptotic
transverse momentum $k_{\rho }$ can acquire a very high value compared to
the low initial transversal momentum $v_{\rho 0}$ due to the collision event:
The lower the initial perpendicular velocity, the higher the final perpendicular
velocity. 1HCRT, which are also born during the first half cycle of the pulse, 
are completely different from the SFA $k_{\rho }=v_{0\rho }$ prediction,
as seen in Fig. \ref{krho-vrho0}c. The limit of zero perpendicular initial velocity
corresponds to a head on collision, so the final perpendicular velocity is high.
The limiting case between 1HCNT and 1HCRT corresponds to trajectories which
start with a given value of $v_{\rho 0}$ ($0.1\lesssim v_{0\rho }\lesssim 0.2$)
and finish with $k_{\rho }=0$, which means that the
electrons move asymptotically parallel to the polarization axis in the
forward direction. This type of trajectories in the border between
rescattered and non-rescattered trajectories is named glory rescattering
trajectories \cite{Xia18}. This is equivalent to the problem of the family of
orbits encountered when particles are emitted in all directions from a point
source in the presence of a Coulomb potential whose center is displaced with
respect to the source \cite{Samengo94}.

On the other hand, none of the
trajectories released during the second half cycle of the pulse suffer
rescattering, i.e., $v_{\perp 0}k_{\perp }>0$ and the corresponding $%
(k_{\rho },v_{\rho 0}$) map is plotted in Fig. \ref{krho-vrho0}d. Due to the
same Coulomb focusing effect in the case of 1HCNT, trajectories born during
the second half cycle, which we call 2HCT, are weakly affected by the
potential of the remaining core and slightly departs from the SFA prediction
(dashed line $k_{\rho }=v_{\rho 0}).$ We see from Fig. \ref{krho-vrho0}d
that 2HCT have a map similar to the SFA $k_{\rho }=v_{0\rho }$ with no
rescattering at all. Dislike 1HCNT, there is no lower limit for the
initial transverse momentum $v_{\rho 0}$ for 2HCT. Therefore, even for low
values of $v_{0\rho }$, the SFA is a good approximation for 2HCT. Summing
up, three different types of trajectories are present in the atomic
ionization process: 1HCNT ionized during the first half cycle that do not suffer
rescattering, 1HCRT ionized during the first half cycle which suffer
rescattering, and 2HCT ionized during the 2HC (which do not suffer
rescattering).

%--------------------------------------------------------------------------
\begin{figure}[tbp]
\centering
\includegraphics[width=0.6\textwidth]{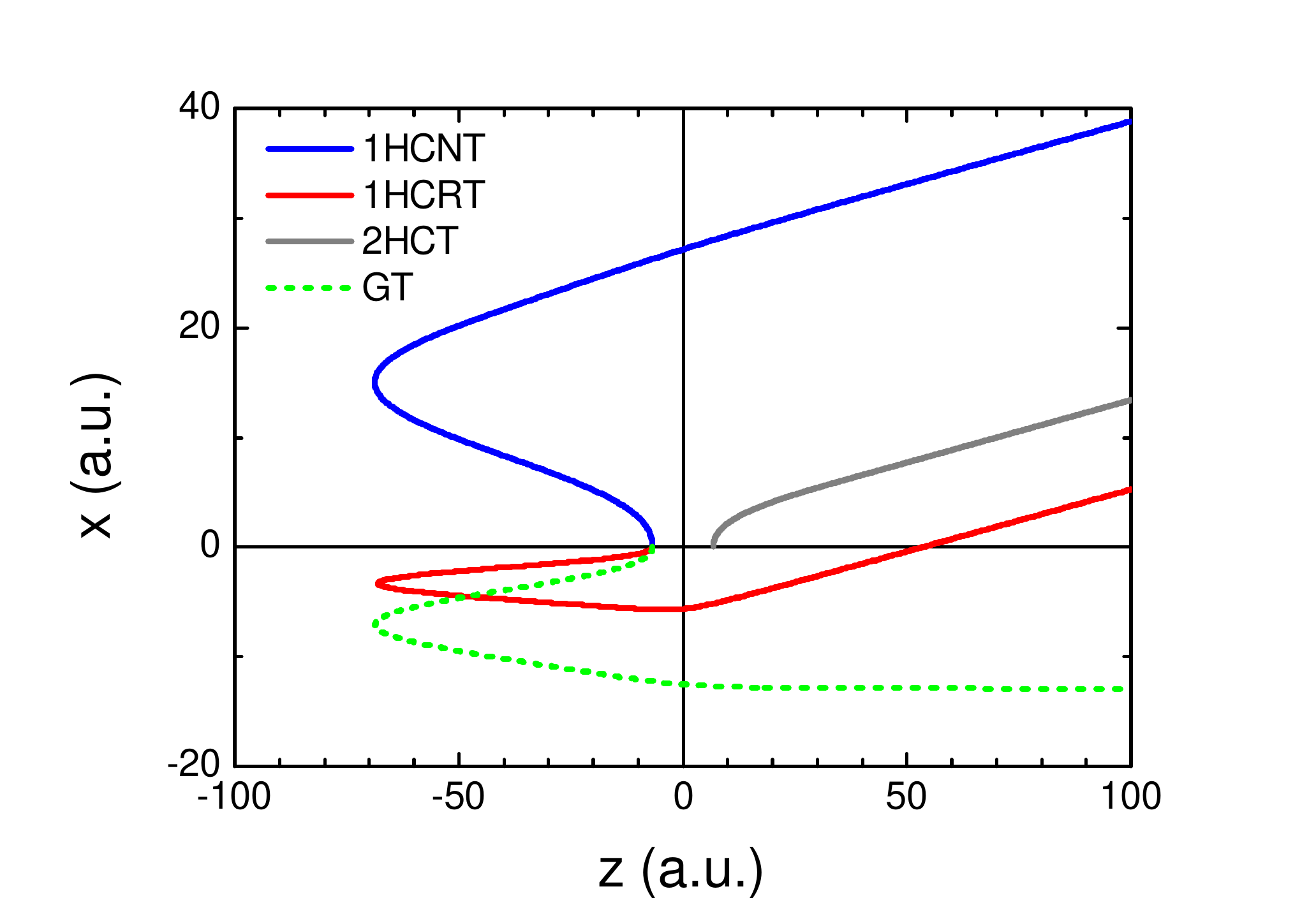}
\caption{Examples of the three different types of trajectories present
in the photoionization process: In blue 1HCNT, in red 1HCRT, and in gray
2HCT. All of these three trajectories have asymptotic momentum $k_z=1.5$ a.u.
and $k_{\bot}=0.17$. Besides, in dotted line the GT 
(transition between 1HCNT and 1HCRT, i.e., $k_{\bot}=0$).
}
\label{trajectories}
\end{figure}
%--------------------------------------------------------------------------

In Fig. \ref{trajectories} we show one example for the three different types of 
electron trajectories 1HCNT, 1HCRT, and 2HCT with the same asymptotic momentum 
$k_z=1.5$ a.u. and $k_{\bot}=0.17$ (corresponding to the first minimum 
in the holographic structure with longitudinal momentum close to maximum emission
according to the SMM). The two trajectories released during the first half cycle
1HCNT and 1HCRT have about the same ionization time (within the statistical uncertainty)
$t_0=28.02$ a.u. This is a general characteristic for all 1HCNT and 1HCRT.
The initial position after the first step (tunneling) depends only on the ionization
time and is $z_0=-\sqrt{I_P/F(t_0)}=-6.76$ a.u. for both types of trajectories.
What makes the difference between the two trajectories 1HCNT and 1HCRT is the initial 
transverse momentum which is $v_{\bot 0}=0.265$ a.u. for 1HCNT and $v_{\bot 0}=-0.0594$ 
for the 1HCRT. We can see an abrupt change of direction in the 1HCRT due to the collision
with the nucleus when $z\simeq 0$, which changes the direction of the transversal velocity.
This collision takes place at $t=149.76$ a.u., which is about $25$ a.u. (~20 \% of an optical
cycle) after the end of the electric field ($2\pi/\omega=125.66$ a.u.).
The trajectory released within the second half cycle 2HCT starts its
trip in the continuum at $t_0=90.69$ a.u., which is close to the SMM prediction that
the sum of the ionization times for 1HCNT and 2HCT is $2\pi/\omega=125.66$ a.u.
The difference stems from the action of the Coulomb potential on the electron trajectories
(departing from the SFA). The initial position is $z_0=-\sqrt{I_P/F(t_0)}=6.77$,
almost the same as the trajectories released during the first half cycle but with the
opposite sign due to the inversion of the potential barrier. Finally, we plot an example
of a GT with the same asymptotic longitudinal momentum $k_z=1.5$ and, by definition,
null asymptotic transversal momentum since the collision is not enough to bend the trajectory
so that the final transverse momentum has opposite sign to the initial transverse velocity,
that in this case is $v_{\bot 0}=-0.126$. There are three different values of the initial
perpendicular velocity $v_{0\rho }$ contributing to the electron yield with a particular
value of the final momentum $k_{\rho}$ corresponding to the three different types of
trajectories.

%--------------------------------------------------------------------------
\begin{figure}[tbp]
\centering
\includegraphics[width=0.9\textwidth]{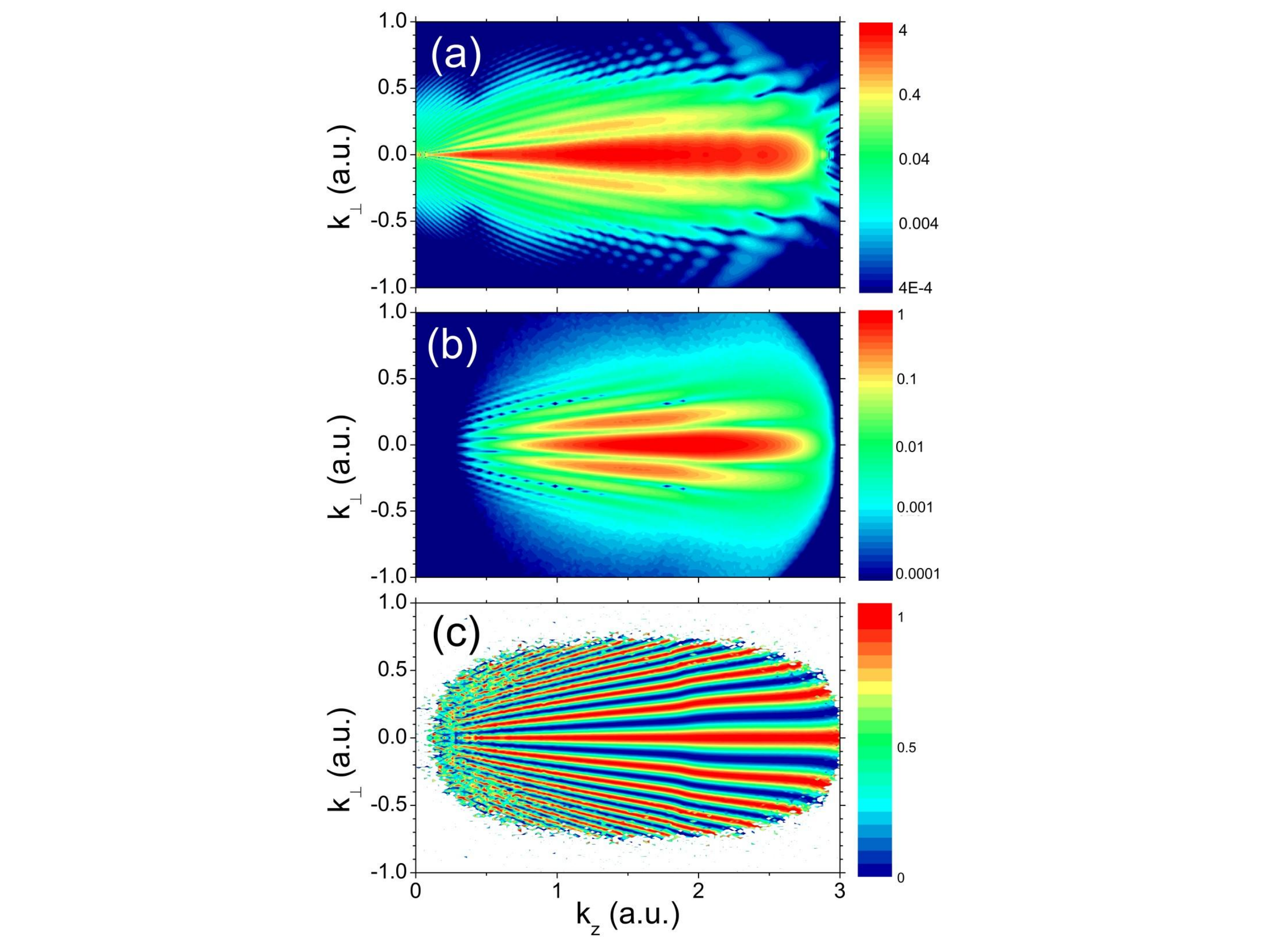}
\caption{Doubly differential momentum distribution of ionization of
atomic hydrogen by the one-cycle sine pulse of Eq. (\ref{field}) when ionization
takes place only during the first half cycle of the pulse (removing ionization
during the second half cycle) within (a) the TDSE and (b) the SCTS. (c) SCTS
holographic interference pattern $\cos ^{2}\left[ \left( \Phi _{\mathrm{1HCRT%
}}-\Phi _{\mathrm{1HCNT}}\right) /2\right] $.
}
\label{holo-mom-dist}
\end{figure}
%--------------------------------------------------------------------------

In order to isolate the holographic interference from the imbroglio of quantum
interference patterns in Fig. \ref{full-mom-dist}a, we artificially switch
off the ionization during the second half cycle (2HC) of the pulse by
projecting the wave function at the middle of the pulse $\Psi (t=\pi /\omega
)$ onto the continuum states dropping out, in this way, the remaining bound
states' populations. The time evolution afterwards ($\pi /\omega <t\leq 2\pi
/\omega $) continues normally allowing recapture and further ionization.
In this way, only the electron yield ionized
during the first half cycle of the pulse and driven by the whole pulse is
considered. The corresponding doubly differential momentum distributions can
be seen in Fig. \ref{holo-mom-dist}a. As it can be clearly observed, the
well-known holographic structure hindered in Fig. \ref{full-mom-dist}a by
other types of interferences comes up. 
This scheme has been recently used for multiple-cycle pulses \cite%
{Xie16,Lopez19,Borbely19}. In a multiple-cycle pulse, the main
lobe centered at $k_{\perp }=0$ flanked by a family of thinner stripes
extends also to the backward ($k_{z}<0$) direction. 
If we exclude ionization during the second half cycle in our
semiclassical calculations, only trajectories 1HCNT in Fig. \ref{krho-vrho0}%
b and 1HCRT in Fig. \ref{krho-vrho0}c contribute to ionization.
The holographic pattern in the doubly differential momentum distribution
in Fig. \ref{holo-mom-dist}b arises as the interference of the two different 
kinds of electron trajectories: rescattering (1HCRT) and non-rescattering 
(1HCNT) trajectories ionized during the first half cycle. From a semiclassical
perspective, electron trajectories of one kind have a certain
accumulated phase [Eq. (\ref{phase})] from the ionization time up to the
final state denoted by a particular momentum $\vec{k}$ in the momentum plane 
$(k_{z},k_{\perp })$ and interfere with the other kind of trajectories
having a different accumulated phase. The similarity of the SCTS (in Fig. %
\ref{holo-mom-dist}a) and TDSE (in Fig. \ref{holo-mom-dist}b) holographic
interference pattern is very good. Considering a holographic interference
nomenclature, 1HCNT plays the role of the reference beam, whereas 1HCRT does
it of the signal beam.

We can enhance the interference pattern calculating the phase of each
electron trajectory and average it within every momentum bin in the
two-dimensional grid for every electron trajectory type, i.e., 
\begin{equation}
<\Phi _{s}>(k_{zi},k_{\perp j})=\sum_{n}\frac{\Phi _{n}(k_{z},k_{\perp })}{%
N_{ij}},
\end{equation}
where the sum extends over all the $N_{ij}$ electron trajectories with final
momentum $k_{zi}-\Delta k_{zi}/2<k_{z}<k_{zi}+\Delta k_{zi}/2,$ and $%
k_{\perp j}-\Delta k_{\perp j}/2<k_{\perp }<k_{\perp j}+\Delta k_{\perp j}/2$
and the grid $(k_{zi},k_{\perp j})$ span the two-dimensional momentum space.
The subscript $s$ denotes the type of electron trajectories, i.e., 1HCRT,
1HCNT, and 2HCT. Then, the interference map is calculated as 
\begin{equation}
\cos ^{2}\left[ \frac{<\Phi _{s}>(k_{zi},k_{\perp j})-<\Phi _{s^{\prime
}}>(k_{zi},k_{\perp j})}{2}\right] .  \label{interf-map}
\end{equation}

In Fig. \ref{holo-mom-dist}c we show the holographic interference map
stemming from the calculation of Eq. (\ref{interf-map}) for $s=$1HCRT, and $%
s^{\prime }=$ 1HCNT. The white color corresponds to regions in the $%
(k_{z},k_{\perp })$ plane with no trajectories of either $s$ or $s^{\prime }$
type. The general shape of the holographic interference pattern in Fig. \ref%
{holo-mom-dist}c shows radial stripes with a moderate jump at $k_{z}\simeq 1$
decreasing slightly this value as $k_{\perp }$ increases. The holographic
map calculated from Eq. (\ref{interf-map}) allows to see the interference
pattern even for momentum regions where the probability distribution is very
low (less than four orders of magnitude lower than the maximum in our case) in
Fig. \ref{holo-mom-dist}c.

%--------------------------------------------------------------------------
\begin{figure}[tbp]
\centering
\includegraphics[width=0.6\textwidth]{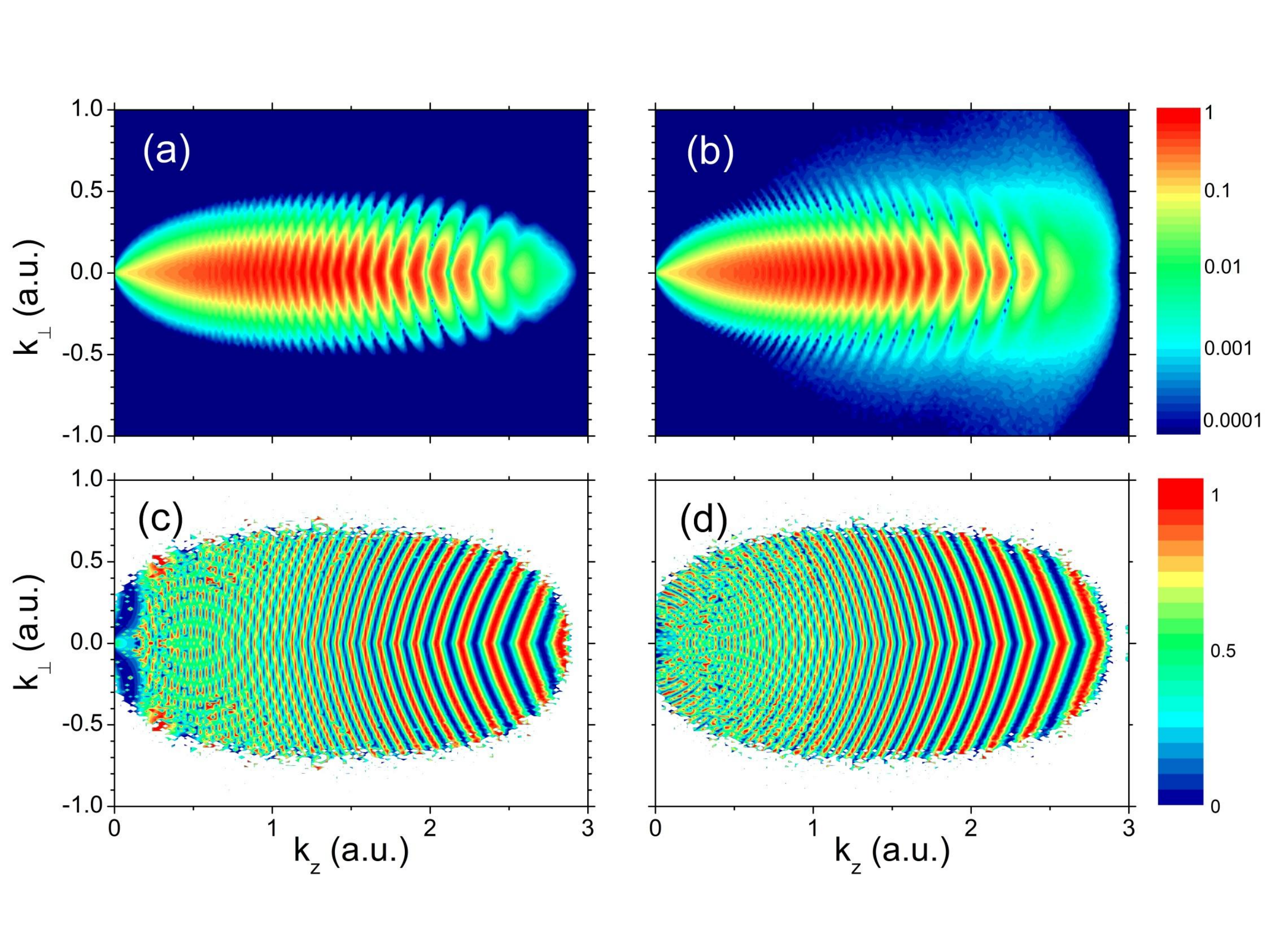}
\caption{(a) SCTS intracycle interference pattern type I considering
only non-rescattering trajectories during the whole pulse doubly
differential momentum distribution of ionization of atomic hydrogen. (b)
SCTS intracycle interference pattern type II considering direct and
rescattered indirect trajectories. (c) Intracycle interference pattern type
I $\cos ^{2}\left[ \left( \Phi _{\mathrm{2HCT}}-\Phi _{\mathrm{1HCNT}%
}\right) /2\right] $. (d) Intracycle interference pattern type II $\cos ^{2}%
\left[ \left( \Phi _{\mathrm{2HCT}}-\Phi _{\mathrm{1HCRT}}\right) /2\right]$.
The laser parameters are the same as Fig. \ref{full-mom-dist}.
}
\label{intra-mom-dist}
\end{figure}
%--------------------------------------------------------------------------

So far, we have analyzed only one type of interference: The holographic
interference. On the other hand, two other types of interference naturally
come up: the interference between 1HCNT and 2HCT, which we name intracycle
type I and is known in the literature as intracycle interference \cite%
{Arbo06b,Arbo12,Arbo10a}, and the interference between 1HCRT and 2HCT, that
we name intracycle type II and are scarcely studied in the literature \cite%
{Maxwell18a,Maxwell18b}. Fig. \ref{intra-mom-dist}a shows the results of the
intracycle interference type I between 1HCNT and 2HCT as a pattern of convex
boomerang-shape stripes centered at the longitudinal momentum axis. The only
difference between the intracycle pattern type I of Fig. \ref{intra-mom-dist}%
a and the previously studied in the literature \cite{Arbo06b,Arbo12,Arbo10a}
is the pointy edge at $k_{\perp }=0$. One can adjudicate the reason of the
pointy edge in Fig. \ref{intra-mom-dist}a to the action of the potential of
the remaining core on the escaping electron, which was neglected in previous
calculations relying on the strong field approximation \cite%
{Arbo06b,Arbo12,Arbo10a}. We corroborate this result by neglecting the
action of the Coulomb potential in our SCTS simulations (not shown). The
intracycle interference type II between 1HCRT and 2HCT in Fig. \ref%
{intra-mom-dist}b has a similar shape of the intracycle interference type I
but the stripes are concave with the pointy edge aiming at the positive 
$k_{z}$ axis. The respective interference maps of Eq. (\ref{interf-map})
are shown in Fig. \ref{intra-mom-dist}c and \ref{intra-mom-dist}d for the
intracycle interferences type I and type II. As in holographic
interference, we see how the interference pattern is enhanced in the
interference maps for the intracycle interferences type I and II. This fact
clearly demonstrates that the stripes in Figs. \ref{intra-mom-dist}a and \ref%
{intra-mom-dist}b stem from the interference of the corresponding types of
trajectories. Some odd non-physical moir\'{e} patterns can be seen for low
parallel momentum because of the plotting method \cite{Dran18}.

%--------------------------------------------------------------------------
\begin{figure}[tbp]
\centering
\includegraphics[width=.7\textwidth]{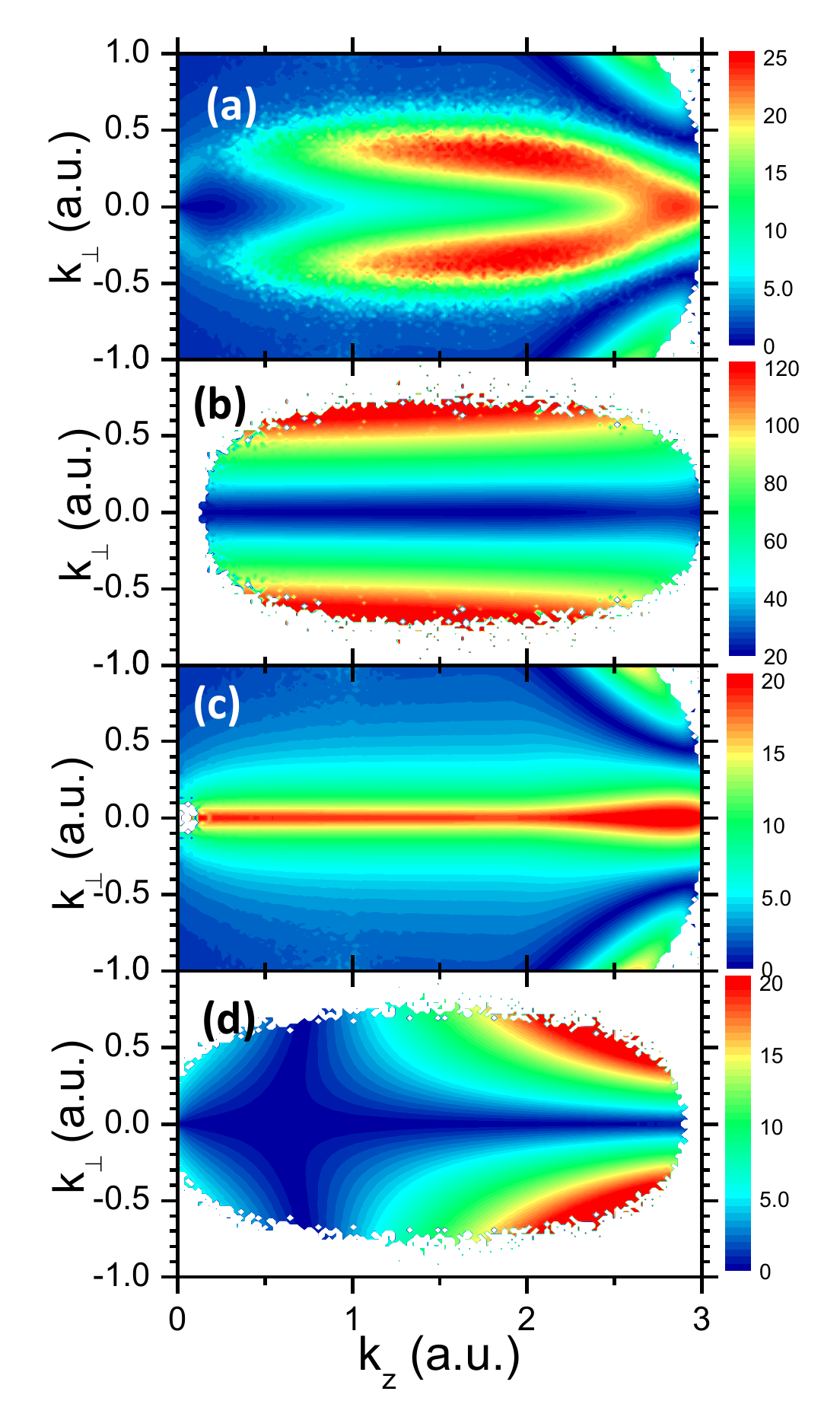}
\caption{Angular momentum as a function of the transversal and
longitudinal momentum. (a) All trajectories, (b) 1HCNT (non-rescattered
trajectories ionized during first half cycle), (c) 1HCRT (rescattered trajectories
ionized during first half cycle), (d) 2HCT (trajectories ionized during the second
half cycle). The laser parameters are the same as in Fig. \ref{full-mom-dist}.
}
\label{ang-mom-map}
\end{figure}
%--------------------------------------------------------------------------

As we have seen, the three different types of electron trajectories present
in the ionization process situate at essentially the same region (forward
emission) in the $(k_{z},k_{\perp })$ momentum plane. Therefore, it is
difficult to identify them without tracing back their time evolution. One way
to discriminate among the different types of trajectories is looking at
their angular momenta. In Fig. \ref{ang-mom-map}a, we show the average
angular momentum 
\begin{equation}
<L>(k_{zi},k_{\perp j})=\sum_{n}\frac{L_{n}(k_{z},k_{\perp })}{N_{i,j}},
\end{equation}
where the sum extends over all the electron trajectories with final momentum 
$k_{zi}-\Delta k_{zi}/2<k_{z}<k_{zi}+\Delta k_{zi}/2,$ and $k_{\perp
j}-\Delta k_{\perp j}/2<k_{\perp }<k_{\perp j}+\Delta k_{\perp j}/2$ and the
grid $(k_{zi},k_{\perp j})$ spans the two-dimensional momentum space. All
the three types of trajectories are present in Fig. \ref{ang-mom-map}a. The
maximum angular momentum found is about $25$ a.u. in a region similar to a
fork bifurcating at the classical edge $k_{z}=3$ a.u. and $k_{\perp }=0$
and aiming backwards.
White color corresponds to regions of the final momentum space with no
trajectories. We have also performed the same calculation for each of the three
types of trajectories separately. In Fig. \ref{ang-mom-map}b we plot the angular
momentum for 1HCNT as a function of the final momentum $(k_{z},k_{\perp }).$
We see that the minimum angular momentum for 1HCNT is at $k_{\perp }=0$ and $%
<L>$ increases from $20$ a.u. with the absolute value of the perpendicular
velocity $k_{\rho }=|k_{\perp }|$ reaching very high values close to $120$
a.u. at the classical boundaries. The angular momentum of 1HCRT in Fig. \ref%
{ang-mom-map}c exhibits a completely different behavior: The maximum
corresponds to $<L>=20$ a.u. and the angular momentum decreases with the
absolute value of the perpendicular velocity $k_{\rho }=|k_{\perp }|.$
Therefore, we can say that 1HCNT and 1HCRT have different values of angular
momentum, coinciding only for $k_{\perp }=0$. In fact, this is the same
result pointed out in the description of Fig. \ref{krho-vrho0}b and \ref%
{krho-vrho0}c because the definition of rescattering and non-rescattering
trajectories mixes when $k_{\rho }=k_{\perp }=0,$ corresponding to the
rescattering glory trajectories described before. Coming back to the
analysis of the angular momentum, we see in Fig. \ref{krho-vrho0}c that 2HCT
have angular momentum with a minimum at $k_{\perp }=0$ but, in contrast to
the trajectories ionized during the 1HC, the minimum value is $<L>=0$.

\begin{figure}[tbp]
\centering
\includegraphics[width=.5\textwidth]{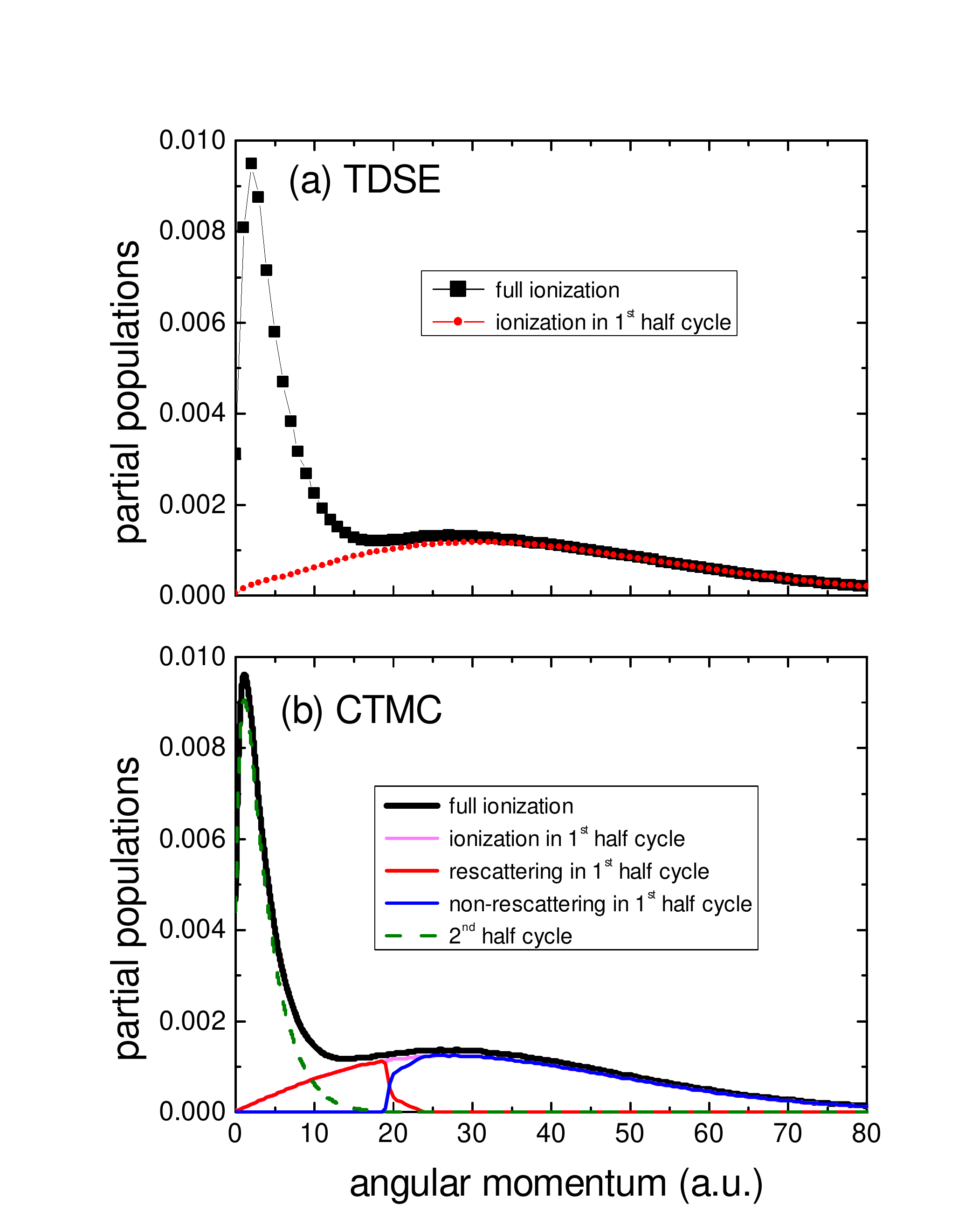}
\caption{Angular momentum distribution for the different kinds of
trajectorie. (a) TDSE calculations and (b) CTMC calculations. The laser
parameters are the same as in Fig. \ref{full-mom-dist}.
}
\label{ang-mom-dist}
\end{figure}

The angular momentum distribution of the electron yield has been calculated
at the end of the pulse, i.e., $t=2\pi /\omega $, which is the same at the
asymptotic detection time $t\rightarrow \infty $ since the angular momentum
is a constant of motion once the laser pulse has been switched off. In Fig. %
\ref{ang-mom-dist}a, we show the quantum angular momentum distribution after
the end of the pulse (calculated within the TDSE). The distribution shows a
sharp peak at very low angular momenta with a broad plateau with maximum at 
$L\simeq 25$ a.u. slowly decreasing up to $L\approx 100$ a.u. In order to
analyze the reason of this shape, we also perform the quantum calculation of
the momentum distribution for ionization during the first half cycle, which
exhibits a very broad distribution with a maximum value at $L\simeq 28$ in 
Fig. \ref{full-mom-dist}. Therefore, we can conclude that the sharp peak at
very low angular momenta stems from ionization during the second half of the pulse.

%--------------------------------------------------------------------------
%--------------------------------------------------------------------------

In order to corroborate this, we calculate the classical angular momentum
distribution [calculated within the CTMC with Eq. (\ref{CTMC-proba})]
for each of the three types of
electron trajectories present in the ionization process. We observe in Fig. %
\ref{ang-mom-dist}b that the sharp peak at low momenta is due almost
exclusively by the contribution of 2HCT, that is, the electron trajectories
born during the second half cycle (which do not suffer rescattering). In
turn, 1HCNT (which do not suffer rescattering either) contributes to the
broad plateau only for angular momentum $L\gtrsim 20$ a.u., as previously
shown in Fig. \ref{krho-vrho0}b. Trajectories released within the first half
cycle ending with lower angular momentum ($L\lesssim 20$ a.u.) suffer rescattering
(1HCRT), contributing to the lower region of the plateau and very little to
the sharp peak at low angular momentum. However, the low energy peak of the
TDSE momentum distribution in Fig. \ref{ang-mom-dist}a is slightly broader
than the corresponding CTMC in Fig. \ref{ang-mom-dist}b. The sum of the CTMC
1HCRT and 1HCNT contributions shown in Fig. \ref{ang-mom-dist}c is very
similar to the quantum distribution with ionization only during the first
half cycle. We can see a very high quantum classical correspondence for the 
angular momentum, even though the resemblance between the quantum and classical
results is not perfect.

%--------------------------------------------------------------------------
\begin{figure}[tbp]
\centering
\includegraphics[width=.5\textwidth]{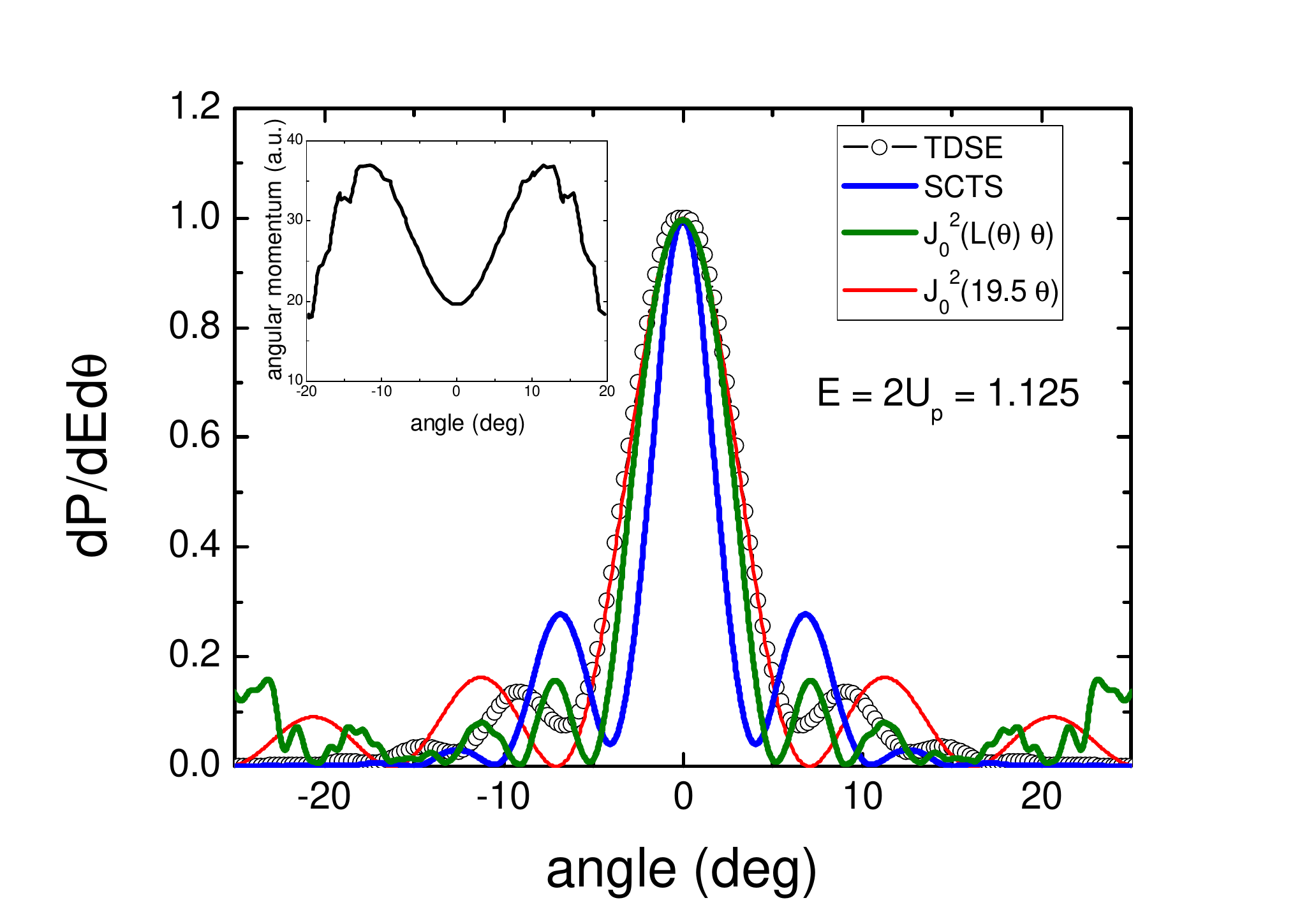}
\caption{Doubly differential angle-energy distribution as a function
of the emission angle for fix energy $E=1.125$ a.u. Calculations are
performed with different methods: dots correspond to TDSE, thick blue line
to SCTS, green line to the square of the Bessel function of the
angle-dependent angular momentum $L(\theta )$ times the angle $\theta ,$ and
thin red line is the same but replacing $L(\theta )$ by $L(0)=19.5.$ The
inset displays the function $L(\theta )$ calculated at $E=1.125$ a.u. The
laser parameters are the same as in Fig. \ref{full-mom-dist}.
All the functions are normalized to unity at forward direction ($\theta=0^{\circ}$)
}
\label{E-ang-dist}
\end{figure}
%--------------------------------------------------------------------------

For the sake of a quantitative comparison between the SCTS and TDSE
calculations, In Fig. \ref{E-ang-dist} we plot the SCTS and TDSE doubly
differential energy-angle distribution as a function of the emission angle
for a fix energy $E=2U_{p}=1.125$ a.u. (corresponding to a momentum $k=1.5$
a.u., close to the maximum of the energy distribution). We observe that the
general shape of the semiclassical and quantum distributions are similar
with a central peak at forward emission ($\theta =0^{\circ }$) and
symmetrical lower peaks at both sides. Despite this qualitative similarity,
several differences are observed. Firstly, the central peak of the SCTS
distribution is narrower than the quantum one. In this sense, the first
minima of the TDSE distribution are at $\theta \simeq \pm 6.5^{\circ },$
whereas the corresponding SCTS ones are situated at $\theta =\pm 4^{\circ }.$
Besides, the position of the first peaks of the TDSE distribution is at $%
\theta \simeq \pm 9^{\circ }$ with a height of $0.135$ relative to the
central peak (normalized in the figure) whereas the corresponding SCTS ones
lie at $\theta \simeq \pm 7^{\circ }$ with height of $0.28.$ According to
Eq. (\ref{GRT}) in Sec. \ref{theory}, due to an interference process of
glory trajectories, the angular distribution (for a fix energy) can be
described by the square of a Bessel function of the first type of the
angular momentum times the angle, i.e., $J_{0}^{2}(L\ \theta )$ (see Ref. 
\cite{Xia18}). Therefore, in the inset of Fig. \ref{E-ang-dist} we show
the SCTS angular momentum $L(\theta )$ as a function of the angle for the
same fix energy $E=1.125$ a.u. The value of the angular momentum at zero
emission angle ($\theta =0^{\circ }$) is $L(0^{\circ })=19.5$ a.u., then it
increases as the emission angle departs from the forward emission up to $%
|\theta |\simeq 12^{\circ }$ where it reaches the maximum angular momentum
and then decreases as the emission angle increases further $|\theta |\simeq
12^{\circ }$. In a thin red line we show that $J_{0}^{2}(L(0^{\circ })\ \theta
)$ follows the TDSE distribution very accurately for low emission angles,
i.e., $|\theta |\lesssim 6^{\circ }$ but then it predicts angles of minima
and maxima higher than the quantum simulation. If we replace the constant
value of $L(0^{\circ })$ by the function $L(\theta )$ into the Bessel
function, the prediction $J_{0}^{2}(L(\theta )\ \theta )$ changes
considerably. It departs from the SCTS at lower angles $|\theta |\simeq
3^{\circ }$ with first maxima at $|\theta |\simeq 7^{\circ },$ very close to
the semiclassical prediction, although higher order peaks set off from the
SCTS.

\section{\label{conclusions}Conclusions}

%---------------------------------------------------

We have studied the interference phenomena in atomic ionization by a
single-cycle laser pulse. We have shown that the SCTS qualitatively reproduces the
quantum results. Within the SCTS model we have identified three
different types of electron trajectories with three different types of
interferences. Non-rescattering trajectories (1HCNT) and rescattering
trajectories (1HCRT) lead to the well-known holographic interference pattern
in the doubly differential momentum distribution. We have shown that one way
to distinguish between these two types of trajectories (1HCNT and 1HCRT)
is through their different final angular momentum.
We have revisited the glory rescattering theory of Ref. \cite{Xia18}
and found that it qualitatively explains the TDSE holographic
interference pattern but some quantitative discrepancies arise.
Moreover, glory trajectories ($k_{\perp} \simeq 0$) are in
the transition between non-rescattering trajectories 1HCNT (where $k_{\perp
}v_{x0}>0$) and rescattering trajectories 1HCRT (where $k_{\perp }v_{x0}<0$%
). For this reason, we have dropped out the name
\textquotedblleft rescattering\textquotedblright\ used in Ref. \cite{Xia18}
and just call them glory trajectories.

Electron trajectories born during the second half cycle 2HCT do not suffer
rescattering and interfere with the other two types of trajectories (released during
the first half cycle). On one hand, the interference between 2HCT and 1HCNT
gives rise to the well-known intracycle interference type I \cite%
{Chirila05,Arbo06a,Arbo06b}. The intracycle type I interference calculated
within the SCTS exhibits a family of convex pointy stripes in the doubly
differential momentum distribution. On the other hand, the interference
between 2HCT and 1HCRT gives rise to the intracycle interference type II as a
family of concave pointy stripes in the momentum distribution. 
Whereas the wedges of the stripes of the interference type I aim to the 
backward direction, those corresponding to interference type II do it forwards.
The sharp wedges at $k_{\perp }=0$ for
both intracycle interferences type I and II are due to the effect of the
Coulomb potential with the escaping electron and is not present in previous
calculations based on the SFA \cite{Chirila05,Arbo06a,Arbo06b}.

Finally, we have shown both quantum mechanically and classically that the very
sharp peak at low angular momentum in the angular momentum distribution
mostly stems from ionization during the second half cycle, whereas ionization
during the first half cycle contributes to the very broad plateau reaching
high values of the angular momentum up to $L\sim 100$.

\begin{acknowledgments}
Work supported by CONICET PIP0386, PICT-2016-0296, PICT-2017-2945, and
PICT-2016-3029 of ANPCyT (Argentina).
\end{acknowledgments}

%---------------------------------------------------
%\bibliographystyle{plain}
\bibliography{biblio-diego-r}

\begin{thebibliography}{79}%
\makeatletter
\providecommand \@ifxundefined [1]{%
 \@ifx{#1\undefined}
}%
\providecommand \@ifnum [1]{%
 \ifnum #1\expandafter \@firstoftwo
 \else \expandafter \@secondoftwo
 \fi
}%
\providecommand \@ifx [1]{%
 \ifx #1\expandafter \@firstoftwo
 \else \expandafter \@secondoftwo
 \fi
}%
\providecommand \natexlab [1]{#1}%
\providecommand \enquote  [1]{``#1''}%
\providecommand \bibnamefont  [1]{#1}%
\providecommand \bibfnamefont [1]{#1}%
\providecommand \citenamefont [1]{#1}%
\providecommand \href@noop [0]{\@secondoftwo}%
\providecommand \href [0]{\begingroup \@sanitize@url \@href}%
\providecommand \@href[1]{\@@startlink{#1}\@@href}%
\providecommand \@@href[1]{\endgroup#1\@@endlink}%
\providecommand \@sanitize@url [0]{\catcode `\\12\catcode `\$12\catcode
  `\&12\catcode `\#12\catcode `\^12\catcode `\_12\catcode `\%12\relax}%
\providecommand \@@startlink[1]{}%
\providecommand \@@endlink[0]{}%
\providecommand \url  [0]{\begingroup\@sanitize@url \@url }%
\providecommand \@url [1]{\endgroup\@href {#1}{\urlprefix }}%
\providecommand \urlprefix  [0]{URL }%
\providecommand \Eprint [0]{\href }%
\providecommand \doibase [0]{http://dx.doi.org/}%
\providecommand \selectlanguage [0]{\@gobble}%
\providecommand \bibinfo  [0]{\@secondoftwo}%
\providecommand \bibfield  [0]{\@secondoftwo}%
\providecommand \translation [1]{[#1]}%
\providecommand \BibitemOpen [0]{}%
\providecommand \bibitemStop [0]{}%
\providecommand \bibitemNoStop [0]{.\EOS\space}%
\providecommand \EOS [0]{\spacefactor3000\relax}%
\providecommand \BibitemShut  [1]{\csname bibitem#1\endcsname}%
\let\auto@bib@innerbib\@empty
%</preamble>
\bibitem [{\citenamefont {{van de Hulst}}(1947)}]{Hulst47}%
  \BibitemOpen
  \bibfield  {author} {\bibinfo {author} {\bibfnamefont {H.~C.}\ \bibnamefont
  {{van de Hulst}}},\ }\href@noop {} {\bibfield  {journal} {\bibinfo  {journal}
  {Journal of the Optical Society of America (1917-1983)}\ }\textbf {\bibinfo
  {volume} {37}},\ \bibinfo {pages} {16} (\bibinfo {year} {1947})}\BibitemShut
  {NoStop}%
\bibitem [{\citenamefont {{Nussenzveig}}(1992)}]{Nussenzveig92}%
  \BibitemOpen
  \bibfield  {author} {\bibinfo {author} {\bibfnamefont {H.~M.}\ \bibnamefont
  {{Nussenzveig}}},\ }\href@noop {} {\emph {\bibinfo {title} {Diffraction
  Effects in Semiclassical Scattering, by H.~M.~Nussenzveig, pp.~252.~ISBN
  0521383188.~Cambridge, UK: Cambridge University Press, July 1992.}}}\
  (\bibinfo {year} {1992})\ p.\ \bibinfo {pages} {252}\BibitemShut {NoStop}%
\bibitem [{\citenamefont {{Berry}}(2015)}]{Berry15}%
  \BibitemOpen
  \bibfield  {author} {\bibinfo {author} {\bibfnamefont {M.~V.}\ \bibnamefont
  {{Berry}}},\ }\href {\doibase 10.1080/00107514.2015.971625} {\bibfield
  {journal} {\bibinfo  {journal} {Contemporary Physics}\ }\textbf {\bibinfo
  {volume} {56}},\ \bibinfo {pages} {2} (\bibinfo {year} {2015})}\BibitemShut
  {NoStop}%
\bibitem [{\citenamefont {{Swenson}}\ \emph {et~al.}(1989)\citenamefont
  {{Swenson}}, \citenamefont {{Havener}}, \citenamefont {{Stolterfoht}},
  \citenamefont {{Sommer}},\ and\ \citenamefont {{Meyer}}}]{Swenson89}%
  \BibitemOpen
  \bibfield  {author} {\bibinfo {author} {\bibfnamefont {J.~K.}\ \bibnamefont
  {{Swenson}}}, \bibinfo {author} {\bibfnamefont {C.~C.}\ \bibnamefont
  {{Havener}}}, \bibinfo {author} {\bibfnamefont {N.}~\bibnamefont
  {{Stolterfoht}}}, \bibinfo {author} {\bibfnamefont {K.}~\bibnamefont
  {{Sommer}}}, \ and\ \bibinfo {author} {\bibfnamefont {F.~W.}\ \bibnamefont
  {{Meyer}}},\ }\href {\doibase 10.1103/PhysRevLett.63.35} {\bibfield
  {journal} {\bibinfo  {journal} {\prl}\ }\textbf {\bibinfo {volume} {63}},\
  \bibinfo {pages} {35} (\bibinfo {year} {1989})}\BibitemShut {NoStop}%
\bibitem [{\citenamefont {{Swenson}}\ \emph {et~al.}(1991)\citenamefont
  {{Swenson}}, \citenamefont {{Burgd{\"o}rfer}}, \citenamefont {{Meyer}},
  \citenamefont {{Havener}}, \citenamefont {{Gregory}},\ and\ \citenamefont
  {{Stolterfoht}}}]{Swenson91}%
  \BibitemOpen
  \bibfield  {author} {\bibinfo {author} {\bibfnamefont {J.~K.}\ \bibnamefont
  {{Swenson}}}, \bibinfo {author} {\bibfnamefont {J.}~\bibnamefont
  {{Burgd{\"o}rfer}}}, \bibinfo {author} {\bibfnamefont {F.~W.}\ \bibnamefont
  {{Meyer}}}, \bibinfo {author} {\bibfnamefont {C.~C.}\ \bibnamefont
  {{Havener}}}, \bibinfo {author} {\bibfnamefont {D.~C.}\ \bibnamefont
  {{Gregory}}}, \ and\ \bibinfo {author} {\bibfnamefont {N.}~\bibnamefont
  {{Stolterfoht}}},\ }\href {\doibase 10.1103/PhysRevLett.66.417} {\bibfield
  {journal} {\bibinfo  {journal} {Physical Review Letters}\ }\textbf {\bibinfo
  {volume} {66}},\ \bibinfo {pages} {417} (\bibinfo {year} {1991})}\BibitemShut
  {NoStop}%
\bibitem [{\citenamefont {{Barrachina}}\ and\ \citenamefont
  {{Macek}}(1989)}]{Barrachina89}%
  \BibitemOpen
  \bibfield  {author} {\bibinfo {author} {\bibfnamefont {R.~O.}\ \bibnamefont
  {{Barrachina}}}\ and\ \bibinfo {author} {\bibfnamefont {J.~H.}\ \bibnamefont
  {{Macek}}},\ }\href {\doibase 10.1088/0953-4075/22/13/019} {\bibfield
  {journal} {\bibinfo  {journal} {Journal of Physics B Atomic Molecular
  Physics}\ }\textbf {\bibinfo {volume} {22}},\ \bibinfo {pages} {2151}
  (\bibinfo {year} {1989})}\BibitemShut {NoStop}%
\bibitem [{\citenamefont {{Samengo}}\ and\ \citenamefont
  {{Barrachina}}(1996)}]{Samengo96}%
  \BibitemOpen
  \bibfield  {author} {\bibinfo {author} {\bibfnamefont {I.}~\bibnamefont
  {{Samengo}}}\ and\ \bibinfo {author} {\bibfnamefont {R.~O.}\ \bibnamefont
  {{Barrachina}}},\ }\href {\doibase 10.1088/0953-4075/29/18/014} {\bibfield
  {journal} {\bibinfo  {journal} {Journal of Physics B Atomic Molecular
  Physics}\ }\textbf {\bibinfo {volume} {29}},\ \bibinfo {pages} {4179}
  (\bibinfo {year} {1996})}\BibitemShut {NoStop}%
\bibitem [{\citenamefont {{Reinhold}}\ \emph {et~al.}(1991)\citenamefont
  {{Reinhold}}, \citenamefont {{Schultz}}, \citenamefont {{Olson}},
  \citenamefont {{Kelbch}}, \citenamefont {{Koch}},\ and\ \citenamefont
  {{Schmidt-B{\"o}cking}}}]{Reinhold91}%
  \BibitemOpen
  \bibfield  {author} {\bibinfo {author} {\bibfnamefont {C.~O.}\ \bibnamefont
  {{Reinhold}}}, \bibinfo {author} {\bibfnamefont {D.~R.}\ \bibnamefont
  {{Schultz}}}, \bibinfo {author} {\bibfnamefont {R.~E.}\ \bibnamefont
  {{Olson}}}, \bibinfo {author} {\bibfnamefont {C.}~\bibnamefont {{Kelbch}}},
  \bibinfo {author} {\bibfnamefont {R.}~\bibnamefont {{Koch}}}, \ and\ \bibinfo
  {author} {\bibfnamefont {H.}~\bibnamefont {{Schmidt-B{\"o}cking}}},\ }\href
  {\doibase 10.1103/PhysRevLett.66.1842} {\bibfield  {journal} {\bibinfo
  {journal} {Physical Review Letters}\ }\textbf {\bibinfo {volume} {66}},\
  \bibinfo {pages} {1842} (\bibinfo {year} {1991})}\BibitemShut {NoStop}%
\bibitem [{\citenamefont {{Samengo}}\ and\ \citenamefont
  {{Barrachina}}(1994)}]{Samengo94}%
  \BibitemOpen
  \bibfield  {author} {\bibinfo {author} {\bibfnamefont {I.}~\bibnamefont
  {{Samengo}}}\ and\ \bibinfo {author} {\bibfnamefont {R.~O.}\ \bibnamefont
  {{Barrachina}}},\ }\href {\doibase 10.1088/0143-0807/15/6/004} {\bibfield
  {journal} {\bibinfo  {journal} {European Journal of Physics}\ }\textbf
  {\bibinfo {volume} {15}},\ \bibinfo {pages} {300} (\bibinfo {year}
  {1994})}\BibitemShut {NoStop}%
\bibitem [{\citenamefont {{Cohen}}\ \emph {et~al.}(2017)\citenamefont
  {{Cohen}}, \citenamefont {{Kalaitzis}}, \citenamefont {{Danakas}},
  \citenamefont {{L{\'e}pine}},\ and\ \citenamefont {{Bordas}}}]{Cohen17}%
  \BibitemOpen
  \bibfield  {author} {\bibinfo {author} {\bibfnamefont {S.}~\bibnamefont
  {{Cohen}}}, \bibinfo {author} {\bibfnamefont {P.}~\bibnamefont
  {{Kalaitzis}}}, \bibinfo {author} {\bibfnamefont {S.}~\bibnamefont
  {{Danakas}}}, \bibinfo {author} {\bibfnamefont {F.}~\bibnamefont
  {{L{\'e}pine}}}, \ and\ \bibinfo {author} {\bibfnamefont {C.}~\bibnamefont
  {{Bordas}}},\ }\href {\doibase 10.1088/1361-6455/aa5c5e} {\bibfield
  {journal} {\bibinfo  {journal} {Journal of Physics B Atomic Molecular
  Physics}\ }\textbf {\bibinfo {volume} {50}},\ \bibinfo {eid} {065002}
  (\bibinfo {year} {2017})}\BibitemShut {NoStop}%
\bibitem [{\citenamefont {{Kelvich}}\ \emph {et~al.}(2016)\citenamefont
  {{Kelvich}}, \citenamefont {{Becker}},\ and\ \citenamefont
  {{Goreslavski}}}]{Kelvich16}%
  \BibitemOpen
  \bibfield  {author} {\bibinfo {author} {\bibfnamefont {S.~A.}\ \bibnamefont
  {{Kelvich}}}, \bibinfo {author} {\bibfnamefont {W.}~\bibnamefont {{Becker}}},
  \ and\ \bibinfo {author} {\bibfnamefont {S.~P.}\ \bibnamefont
  {{Goreslavski}}},\ }\href {\doibase 10.1103/PhysRevA.93.033411} {\bibfield
  {journal} {\bibinfo  {journal} {\pra}\ }\textbf {\bibinfo {volume} {93}},\
  \bibinfo {eid} {033411} (\bibinfo {year} {2016})}\BibitemShut {NoStop}%
\bibitem [{\citenamefont {{Kelvich}}\ \emph {et~al.}(2017)\citenamefont
  {{Kelvich}}, \citenamefont {{Becker}},\ and\ \citenamefont
  {{Goreslavski}}}]{Kelvich17}%
  \BibitemOpen
  \bibfield  {author} {\bibinfo {author} {\bibfnamefont {S.~A.}\ \bibnamefont
  {{Kelvich}}}, \bibinfo {author} {\bibfnamefont {W.}~\bibnamefont {{Becker}}},
  \ and\ \bibinfo {author} {\bibfnamefont {S.~P.}\ \bibnamefont
  {{Goreslavski}}},\ }\href {\doibase 10.1103/PhysRevA.96.023427} {\bibfield
  {journal} {\bibinfo  {journal} {\pra}\ }\textbf {\bibinfo {volume} {96}},\
  \bibinfo {eid} {023427} (\bibinfo {year} {2017})}\BibitemShut {NoStop}%
\bibitem [{\citenamefont {{Paulus}}\ \emph
  {et~al.}(1994{\natexlab{a}})\citenamefont {{Paulus}}, \citenamefont
  {{Nicklich}},\ and\ \citenamefont {{Walther}}}]{Paulus94a}%
  \BibitemOpen
  \bibfield  {author} {\bibinfo {author} {\bibfnamefont {G.~G.}\ \bibnamefont
  {{Paulus}}}, \bibinfo {author} {\bibfnamefont {W.}~\bibnamefont
  {{Nicklich}}}, \ and\ \bibinfo {author} {\bibfnamefont {H.}~\bibnamefont
  {{Walther}}},\ }\href {\doibase 10.1209/0295-5075/27/4/003} {\bibfield
  {journal} {\bibinfo  {journal} {EPL (Europhysics Letters)}\ }\textbf
  {\bibinfo {volume} {27}},\ \bibinfo {pages} {267} (\bibinfo {year}
  {1994}{\natexlab{a}})}\BibitemShut {NoStop}%
\bibitem [{\citenamefont {{Paulus}}\ \emph
  {et~al.}(1994{\natexlab{b}})\citenamefont {{Paulus}}, \citenamefont
  {{Becker}}, \citenamefont {{Nicklich}},\ and\ \citenamefont
  {{Walther}}}]{Paulus94b}%
  \BibitemOpen
  \bibfield  {author} {\bibinfo {author} {\bibfnamefont {G.~G.}\ \bibnamefont
  {{Paulus}}}, \bibinfo {author} {\bibfnamefont {W.}~\bibnamefont {{Becker}}},
  \bibinfo {author} {\bibfnamefont {W.}~\bibnamefont {{Nicklich}}}, \ and\
  \bibinfo {author} {\bibfnamefont {H.}~\bibnamefont {{Walther}}},\ }\href
  {\doibase 10.1088/0953-4075/27/21/003} {\bibfield  {journal} {\bibinfo
  {journal} {Journal of Physics B Atomic Molecular Physics}\ }\textbf {\bibinfo
  {volume} {27}},\ \bibinfo {pages} {L703} (\bibinfo {year}
  {1994}{\natexlab{b}})}\BibitemShut {NoStop}%
\bibitem [{\citenamefont {{Becker}}\ \emph {et~al.}(2014)\citenamefont
  {{Becker}}, \citenamefont {{Goreslavski}}, \citenamefont {{Milo{\v
  s}evi{\'c}}},\ and\ \citenamefont {{Paulus}}}]{Becker14}%
  \BibitemOpen
  \bibfield  {author} {\bibinfo {author} {\bibfnamefont {W.}~\bibnamefont
  {{Becker}}}, \bibinfo {author} {\bibfnamefont {S.~P.}\ \bibnamefont
  {{Goreslavski}}}, \bibinfo {author} {\bibfnamefont {D.~B.}\ \bibnamefont
  {{Milo{\v s}evi{\'c}}}}, \ and\ \bibinfo {author} {\bibfnamefont {G.~G.}\
  \bibnamefont {{Paulus}}},\ }\href {\doibase 10.1088/0953-4075/47/20/204022}
  {\bibfield  {journal} {\bibinfo  {journal} {Journal of Physics B Atomic
  Molecular Physics}\ }\textbf {\bibinfo {volume} {47}},\ \bibinfo {eid}
  {204022} (\bibinfo {year} {2014})}\BibitemShut {NoStop}%
\bibitem [{\citenamefont {{L'huillier}}\ \emph {et~al.}(1993)\citenamefont
  {{L'huillier}}, \citenamefont {{Lewenstein}}, \citenamefont {{Sali{\`e}res}},
  \citenamefont {{Balcou}}, \citenamefont {{Ivanov}}, \citenamefont
  {{Larsson}},\ and\ \citenamefont {{Wahlstr{\"o}m}}}]{Luillier93}%
  \BibitemOpen
  \bibfield  {author} {\bibinfo {author} {\bibfnamefont {A.}~\bibnamefont
  {{L'huillier}}}, \bibinfo {author} {\bibfnamefont {M.}~\bibnamefont
  {{Lewenstein}}}, \bibinfo {author} {\bibfnamefont {P.}~\bibnamefont
  {{Sali{\`e}res}}}, \bibinfo {author} {\bibfnamefont {P.}~\bibnamefont
  {{Balcou}}}, \bibinfo {author} {\bibfnamefont {M.~Y.}\ \bibnamefont
  {{Ivanov}}}, \bibinfo {author} {\bibfnamefont {J.}~\bibnamefont {{Larsson}}},
  \ and\ \bibinfo {author} {\bibfnamefont {C.~G.}\ \bibnamefont
  {{Wahlstr{\"o}m}}},\ }\href {\doibase 10.1103/PhysRevA.48.R3433} {\bibfield
  {journal} {\bibinfo  {journal} {Phys. Rev. A}\ }\textbf {\bibinfo {volume}
  {48}},\ \bibinfo {pages} {R3433} (\bibinfo {year} {1993})}\BibitemShut
  {NoStop}%
\bibitem [{\citenamefont {{Paulus}}\ \emph
  {et~al.}(1994{\natexlab{c}})\citenamefont {{Paulus}}, \citenamefont
  {{Becker}}, \citenamefont {{Nicklich}},\ and\ \citenamefont
  {{Walther}}}]{Paulus94}%
  \BibitemOpen
  \bibfield  {author} {\bibinfo {author} {\bibfnamefont {G.~G.}\ \bibnamefont
  {{Paulus}}}, \bibinfo {author} {\bibfnamefont {W.}~\bibnamefont {{Becker}}},
  \bibinfo {author} {\bibfnamefont {W.}~\bibnamefont {{Nicklich}}}, \ and\
  \bibinfo {author} {\bibfnamefont {H.}~\bibnamefont {{Walther}}},\ }\href
  {\doibase 10.1088/0953-4075/27/21/003} {\bibfield  {journal} {\bibinfo
  {journal} {Journal of Physics B Atomic Molecular Physics}\ }\textbf {\bibinfo
  {volume} {27}},\ \bibinfo {pages} {L703} (\bibinfo {year}
  {1994}{\natexlab{c}})}\BibitemShut {NoStop}%
\bibitem [{\citenamefont {{Paulus}}\ \emph {et~al.}(1995)\citenamefont
  {{Paulus}}, \citenamefont {{Becker}},\ and\ \citenamefont
  {{Walther}}}]{Paulus95}%
  \BibitemOpen
  \bibfield  {author} {\bibinfo {author} {\bibfnamefont {G.~G.}\ \bibnamefont
  {{Paulus}}}, \bibinfo {author} {\bibfnamefont {W.}~\bibnamefont {{Becker}}},
  \ and\ \bibinfo {author} {\bibfnamefont {H.}~\bibnamefont {{Walther}}},\
  }\href {\doibase 10.1103/PhysRevA.52.4043} {\bibfield  {journal} {\bibinfo
  {journal} {Phys. Rev. A}\ }\textbf {\bibinfo {volume} {52}},\ \bibinfo
  {pages} {4043} (\bibinfo {year} {1995})}\BibitemShut {NoStop}%
\bibitem [{\citenamefont {{Becker}}\ \emph {et~al.}(2002)\citenamefont
  {{Becker}}, \citenamefont {{Grasbon}}, \citenamefont {{Kopold}},
  \citenamefont {{Milo{\v s}evi{\'c}}}, \citenamefont {{Paulus}},\ and\
  \citenamefont {{Walther}}}]{Becker02}%
  \BibitemOpen
  \bibfield  {author} {\bibinfo {author} {\bibfnamefont {W.}~\bibnamefont
  {{Becker}}}, \bibinfo {author} {\bibfnamefont {F.}~\bibnamefont {{Grasbon}}},
  \bibinfo {author} {\bibfnamefont {R.}~\bibnamefont {{Kopold}}}, \bibinfo
  {author} {\bibfnamefont {D.~B.}\ \bibnamefont {{Milo{\v s}evi{\'c}}}},
  \bibinfo {author} {\bibfnamefont {G.~G.}\ \bibnamefont {{Paulus}}}, \ and\
  \bibinfo {author} {\bibfnamefont {H.}~\bibnamefont {{Walther}}},\ }\href
  {\doibase 10.1016/S1049-250X(02)80006-4} {\bibfield  {journal} {\bibinfo
  {journal} {Advances in Atomic Molecular and Optical Physics}\ }\textbf
  {\bibinfo {volume} {48}},\ \bibinfo {pages} {35} (\bibinfo {year}
  {2002})}\BibitemShut {NoStop}%
\bibitem [{\citenamefont {{Su{\'a}rez}}\ \emph {et~al.}(2015)\citenamefont
  {{Su{\'a}rez}}, \citenamefont {{Chac{\'o}n}}, \citenamefont {{Ciappina}},
  \citenamefont {{Biegert}},\ and\ \citenamefont {{Lewenstein}}}]{Suarez15}%
  \BibitemOpen
  \bibfield  {author} {\bibinfo {author} {\bibfnamefont {N.}~\bibnamefont
  {{Su{\'a}rez}}}, \bibinfo {author} {\bibfnamefont {A.}~\bibnamefont
  {{Chac{\'o}n}}}, \bibinfo {author} {\bibfnamefont {M.~F.}\ \bibnamefont
  {{Ciappina}}}, \bibinfo {author} {\bibfnamefont {J.}~\bibnamefont
  {{Biegert}}}, \ and\ \bibinfo {author} {\bibfnamefont {M.}~\bibnamefont
  {{Lewenstein}}},\ }\href {\doibase 10.1103/PhysRevA.92.063421} {\bibfield
  {journal} {\bibinfo  {journal} {Phys. Rev. A}\ }\textbf {\bibinfo {volume}
  {92}},\ \bibinfo {eid} {063421} (\bibinfo {year} {2015})},\ \Eprint
  {http://arxiv.org/abs/1509.01929} {arXiv:1509.01929 [physics.atom-ph]}
  \BibitemShut {NoStop}%
\bibitem [{\citenamefont {{Dan{\v{e}}k}}\ \emph
  {et~al.}(2018{\natexlab{a}})\citenamefont {{Dan{\v{e}}k}}, \citenamefont
  {{Hatsagortsyan}},\ and\ \citenamefont {{Keitel}}}]{Danek18a}%
  \BibitemOpen
  \bibfield  {author} {\bibinfo {author} {\bibfnamefont {J.}~\bibnamefont
  {{Dan{\v{e}}k}}}, \bibinfo {author} {\bibfnamefont {K.~Z.}\ \bibnamefont
  {{Hatsagortsyan}}}, \ and\ \bibinfo {author} {\bibfnamefont {C.~H.}\
  \bibnamefont {{Keitel}}},\ }\href {\doibase 10.1103/PhysRevA.97.063409}
  {\bibfield  {journal} {\bibinfo  {journal} {\pra}\ }\textbf {\bibinfo
  {volume} {97}},\ \bibinfo {eid} {063409} (\bibinfo {year}
  {2018}{\natexlab{a}})},\ \Eprint {http://arxiv.org/abs/1707.06921}
  {arXiv:1707.06921 [physics.atom-ph]} \BibitemShut {NoStop}%
\bibitem [{\citenamefont {{Dan{\v{e}}k}}\ \emph
  {et~al.}(2018{\natexlab{b}})\citenamefont {{Dan{\v{e}}k}}, \citenamefont
  {{Hatsagortsyan}},\ and\ \citenamefont {{Keitel}}}]{Danek18b}%
  \BibitemOpen
  \bibfield  {author} {\bibinfo {author} {\bibfnamefont {J.}~\bibnamefont
  {{Dan{\v{e}}k}}}, \bibinfo {author} {\bibfnamefont {K.~Z.}\ \bibnamefont
  {{Hatsagortsyan}}}, \ and\ \bibinfo {author} {\bibfnamefont {C.~H.}\
  \bibnamefont {{Keitel}}},\ }\href {\doibase 10.1103/PhysRevA.97.063410}
  {\bibfield  {journal} {\bibinfo  {journal} {\pra}\ }\textbf {\bibinfo
  {volume} {97}},\ \bibinfo {eid} {063410} (\bibinfo {year}
  {2018}{\natexlab{b}})}\BibitemShut {NoStop}%
\bibitem [{\citenamefont {Lewenstein}\ \emph {et~al.}(1995)\citenamefont
  {Lewenstein}, \citenamefont {Kulander}, \citenamefont {Schafer},\ and\
  \citenamefont {Bucksbaum}}]{Lewenstein95}%
  \BibitemOpen
  \bibfield  {author} {\bibinfo {author} {\bibfnamefont {M.}~\bibnamefont
  {Lewenstein}}, \bibinfo {author} {\bibfnamefont {K.~C.}\ \bibnamefont
  {Kulander}}, \bibinfo {author} {\bibfnamefont {K.~J.}\ \bibnamefont
  {Schafer}}, \ and\ \bibinfo {author} {\bibfnamefont {P.~H.}\ \bibnamefont
  {Bucksbaum}},\ }\href {\doibase 10.1103/PhysRevA.51.1495} {\bibfield
  {journal} {\bibinfo  {journal} {Phys. Rev. A}\ }\textbf {\bibinfo {volume}
  {51}},\ \bibinfo {pages} {1495} (\bibinfo {year} {1995})}\BibitemShut
  {NoStop}%
\bibitem [{\citenamefont {{Liu}}(2014)}]{Liu14}%
  \BibitemOpen
  \bibfield  {author} {\bibinfo {author} {\bibfnamefont {J.}~\bibnamefont
  {{Liu}}},\ }\href {\doibase 10.1007/978-3-642-40549-5} {\emph {\bibinfo
  {title} {{Classical Trajectory Perspective of Atomic Ionization in Strong
  Laser Fields}}}}\ (\bibinfo {year} {2014})\BibitemShut {NoStop}%
\bibitem [{\citenamefont {{Gribakin}}\ and\ \citenamefont
  {{Kuchiev}}(1997)}]{Gribakin97}%
  \BibitemOpen
  \bibfield  {author} {\bibinfo {author} {\bibfnamefont {G.~F.}\ \bibnamefont
  {{Gribakin}}}\ and\ \bibinfo {author} {\bibfnamefont {M.~Y.}\ \bibnamefont
  {{Kuchiev}}},\ }\href {\doibase 10.1103/PhysRevA.55.3760} {\bibfield
  {journal} {\bibinfo  {journal} {\pra}\ }\textbf {\bibinfo {volume} {55}},\
  \bibinfo {pages} {3760} (\bibinfo {year} {1997})},\ \Eprint
  {http://arxiv.org/abs/physics/9702002} {physics/9702002} \BibitemShut
  {NoStop}%
\bibitem [{\citenamefont {Chirila}\ and\ \citenamefont
  {Potvliege}(2005)}]{Chirila05}%
  \BibitemOpen
  \bibfield  {author} {\bibinfo {author} {\bibfnamefont {C.~C.}\ \bibnamefont
  {Chirila}}\ and\ \bibinfo {author} {\bibfnamefont {R.~M.}\ \bibnamefont
  {Potvliege}},\ }\href {\doibase 10.1103/PhysRevA.71.021402} {\bibfield
  {journal} {\bibinfo  {journal} {Phys. Rev. A}\ }\textbf {\bibinfo {volume}
  {71}},\ \bibinfo {pages} {021402} (\bibinfo {year} {2005})}\BibitemShut
  {NoStop}%
\bibitem [{\citenamefont {Lindner}\ \emph {et~al.}(2005)\citenamefont
  {Lindner}, \citenamefont {Sch\"atzel}, \citenamefont {Walther}, \citenamefont
  {Baltu\ifmmode~\check{s}\else \v{s}\fi{}ka}, \citenamefont {Goulielmakis},
  \citenamefont {Krausz}, \citenamefont {Milo\ifmmode \check{s}\else
  \v{s}\fi{}evi\ifmmode~\acute{c}\else \'{c}\fi{}}, \citenamefont {Bauer},
  \citenamefont {Becker},\ and\ \citenamefont {Paulus}}]{Lindner05}%
  \BibitemOpen
  \bibfield  {author} {\bibinfo {author} {\bibfnamefont {F.}~\bibnamefont
  {Lindner}}, \bibinfo {author} {\bibfnamefont {M.~G.}\ \bibnamefont
  {Sch\"atzel}}, \bibinfo {author} {\bibfnamefont {H.}~\bibnamefont {Walther}},
  \bibinfo {author} {\bibfnamefont {A.}~\bibnamefont
  {Baltu\ifmmode~\check{s}\else \v{s}\fi{}ka}}, \bibinfo {author}
  {\bibfnamefont {E.}~\bibnamefont {Goulielmakis}}, \bibinfo {author}
  {\bibfnamefont {F.}~\bibnamefont {Krausz}}, \bibinfo {author} {\bibfnamefont
  {D.~B.}\ \bibnamefont {Milo\ifmmode \check{s}\else
  \v{s}\fi{}evi\ifmmode~\acute{c}\else \'{c}\fi{}}}, \bibinfo {author}
  {\bibfnamefont {D.}~\bibnamefont {Bauer}}, \bibinfo {author} {\bibfnamefont
  {W.}~\bibnamefont {Becker}}, \ and\ \bibinfo {author} {\bibfnamefont {G.~G.}\
  \bibnamefont {Paulus}},\ }\href {\doibase 10.1103/PhysRevLett.95.040401}
  {\bibfield  {journal} {\bibinfo  {journal} {Phys. Rev. Lett.}\ }\textbf
  {\bibinfo {volume} {95}},\ \bibinfo {pages} {040401} (\bibinfo {year}
  {2005})}\BibitemShut {NoStop}%
\bibitem [{\citenamefont {{Gopal}}\ \emph {et~al.}(2009)\citenamefont
  {{Gopal}}, \citenamefont {{Simeonidis}}, \citenamefont {{Moshammer}},
  \citenamefont {{Ergler}}, \citenamefont {{D{\"u}rr}}, \citenamefont
  {{Kurka}}, \citenamefont {{K{\"u}hnel}}, \citenamefont {{Tschuch}},
  \citenamefont {{Schr{\"o}ter}}, \citenamefont {{Bauer}}, \citenamefont
  {{Ullrich}}, \citenamefont {{Rudenko}}, \citenamefont {{Herrwerth}},
  \citenamefont {{Uphues}}, \citenamefont {{Schultze}}, \citenamefont
  {{Goulielmakis}}, \citenamefont {{Uiberacker}}, \citenamefont {{Lezius}},\
  and\ \citenamefont {{Kling}}}]{Gopal09}%
  \BibitemOpen
  \bibfield  {author} {\bibinfo {author} {\bibfnamefont {R.}~\bibnamefont
  {{Gopal}}}, \bibinfo {author} {\bibfnamefont {K.}~\bibnamefont
  {{Simeonidis}}}, \bibinfo {author} {\bibfnamefont {R.}~\bibnamefont
  {{Moshammer}}}, \bibinfo {author} {\bibfnamefont {T.}~\bibnamefont
  {{Ergler}}}, \bibinfo {author} {\bibfnamefont {M.}~\bibnamefont
  {{D{\"u}rr}}}, \bibinfo {author} {\bibfnamefont {M.}~\bibnamefont {{Kurka}}},
  \bibinfo {author} {\bibfnamefont {K.-U.}\ \bibnamefont {{K{\"u}hnel}}},
  \bibinfo {author} {\bibfnamefont {S.}~\bibnamefont {{Tschuch}}}, \bibinfo
  {author} {\bibfnamefont {C.-D.}\ \bibnamefont {{Schr{\"o}ter}}}, \bibinfo
  {author} {\bibfnamefont {D.}~\bibnamefont {{Bauer}}}, \bibinfo {author}
  {\bibfnamefont {J.}~\bibnamefont {{Ullrich}}}, \bibinfo {author}
  {\bibfnamefont {A.}~\bibnamefont {{Rudenko}}}, \bibinfo {author}
  {\bibfnamefont {O.}~\bibnamefont {{Herrwerth}}}, \bibinfo {author}
  {\bibfnamefont {T.}~\bibnamefont {{Uphues}}}, \bibinfo {author}
  {\bibfnamefont {M.}~\bibnamefont {{Schultze}}}, \bibinfo {author}
  {\bibfnamefont {E.}~\bibnamefont {{Goulielmakis}}}, \bibinfo {author}
  {\bibfnamefont {M.}~\bibnamefont {{Uiberacker}}}, \bibinfo {author}
  {\bibfnamefont {M.}~\bibnamefont {{Lezius}}}, \ and\ \bibinfo {author}
  {\bibfnamefont {M.~F.}\ \bibnamefont {{Kling}}},\ }\href {\doibase
  10.1103/PhysRevLett.103.053001} {\bibfield  {journal} {\bibinfo  {journal}
  {Physical Review Letters}\ }\textbf {\bibinfo {volume} {103}},\ \bibinfo
  {eid} {053001} (\bibinfo {year} {2009})}\BibitemShut {NoStop}%
\bibitem [{\citenamefont {Arb\'o}\ \emph
  {et~al.}(2006{\natexlab{a}})\citenamefont {Arb\'o}, \citenamefont {Persson},\
  and\ \citenamefont {Burgd\"orfer}}]{Arbo06b}%
  \BibitemOpen
  \bibfield  {author} {\bibinfo {author} {\bibfnamefont {D.~G.}\ \bibnamefont
  {Arb\'o}}, \bibinfo {author} {\bibfnamefont {E.}~\bibnamefont {Persson}}, \
  and\ \bibinfo {author} {\bibfnamefont {J.}~\bibnamefont {Burgd\"orfer}},\
  }\href {\doibase 10.1103/PhysRevA.74.063407} {\bibfield  {journal} {\bibinfo
  {journal} {Phys. Rev. A}\ }\textbf {\bibinfo {volume} {74}},\ \bibinfo
  {pages} {063407} (\bibinfo {year} {2006}{\natexlab{a}})}\BibitemShut
  {NoStop}%
\bibitem [{\citenamefont {Arb\'o}\ \emph
  {et~al.}(2010{\natexlab{a}})\citenamefont {Arb\'o}, \citenamefont {Ishikawa},
  \citenamefont {Schiessl}, \citenamefont {Persson},\ and\ \citenamefont
  {Burgd\"orfer}}]{Arbo10a}%
  \BibitemOpen
  \bibfield  {author} {\bibinfo {author} {\bibfnamefont {D.~G.}\ \bibnamefont
  {Arb\'o}}, \bibinfo {author} {\bibfnamefont {K.~L.}\ \bibnamefont
  {Ishikawa}}, \bibinfo {author} {\bibfnamefont {K.}~\bibnamefont {Schiessl}},
  \bibinfo {author} {\bibfnamefont {E.}~\bibnamefont {Persson}}, \ and\
  \bibinfo {author} {\bibfnamefont {J.}~\bibnamefont {Burgd\"orfer}},\ }\href
  {\doibase 10.1103/PhysRevA.81.021403} {\bibfield  {journal} {\bibinfo
  {journal} {Phys. Rev. A}\ }\textbf {\bibinfo {volume} {81}},\ \bibinfo
  {pages} {021403} (\bibinfo {year} {2010}{\natexlab{a}})}\BibitemShut
  {NoStop}%
\bibitem [{\citenamefont {Arb\'o}\ \emph
  {et~al.}(2010{\natexlab{b}})\citenamefont {Arb\'o}, \citenamefont {Ishikawa},
  \citenamefont {Schiessl}, \citenamefont {Persson},\ and\ \citenamefont
  {Burgd\"orfer}}]{Arbo10b}%
  \BibitemOpen
  \bibfield  {author} {\bibinfo {author} {\bibfnamefont {D.~G.}\ \bibnamefont
  {Arb\'o}}, \bibinfo {author} {\bibfnamefont {K.~L.}\ \bibnamefont
  {Ishikawa}}, \bibinfo {author} {\bibfnamefont {K.}~\bibnamefont {Schiessl}},
  \bibinfo {author} {\bibfnamefont {E.}~\bibnamefont {Persson}}, \ and\
  \bibinfo {author} {\bibfnamefont {J.}~\bibnamefont {Burgd\"orfer}},\ }\href
  {\doibase 10.1103/PhysRevA.82.043426} {\bibfield  {journal} {\bibinfo
  {journal} {Phys. Rev. A}\ }\textbf {\bibinfo {volume} {82}},\ \bibinfo
  {pages} {043426} (\bibinfo {year} {2010}{\natexlab{b}})}\BibitemShut
  {NoStop}%
\bibitem [{\citenamefont {{Arb{\'o}}}\ \emph {et~al.}(2012)\citenamefont
  {{Arb{\'o}}}, \citenamefont {{Ishikawa}}, \citenamefont {{Persson}},\ and\
  \citenamefont {{Burgd{\"o}rfer}}}]{Arbo12}%
  \BibitemOpen
  \bibfield  {author} {\bibinfo {author} {\bibfnamefont {D.~G.}\ \bibnamefont
  {{Arb{\'o}}}}, \bibinfo {author} {\bibfnamefont {K.~L.}\ \bibnamefont
  {{Ishikawa}}}, \bibinfo {author} {\bibfnamefont {E.}~\bibnamefont
  {{Persson}}}, \ and\ \bibinfo {author} {\bibfnamefont {J.}~\bibnamefont
  {{Burgd{\"o}rfer}}},\ }\href {\doibase 10.1016/j.nimb.2011.10.030} {\bibfield
   {journal} {\bibinfo  {journal} {Nuclear Instruments and Methods in Physics
  Research B}\ }\textbf {\bibinfo {volume} {279}},\ \bibinfo {pages} {24}
  (\bibinfo {year} {2012})}\BibitemShut {NoStop}%
\bibitem [{\citenamefont {{Yang}}\ \emph {et~al.}(2016)\citenamefont {{Yang}},
  \citenamefont {{Zhang}}, \citenamefont {{Lin}}, \citenamefont {{Xu}},
  \citenamefont {{Sheng}}, \citenamefont {{Song}}, \citenamefont {{Hu}},\ and\
  \citenamefont {{Chen}}}]{Yang16}%
  \BibitemOpen
  \bibfield  {author} {\bibinfo {author} {\bibfnamefont {W.}~\bibnamefont
  {{Yang}}}, \bibinfo {author} {\bibfnamefont {H.}~\bibnamefont {{Zhang}}},
  \bibinfo {author} {\bibfnamefont {C.}~\bibnamefont {{Lin}}}, \bibinfo
  {author} {\bibfnamefont {J.}~\bibnamefont {{Xu}}}, \bibinfo {author}
  {\bibfnamefont {Z.}~\bibnamefont {{Sheng}}}, \bibinfo {author} {\bibfnamefont
  {X.}~\bibnamefont {{Song}}}, \bibinfo {author} {\bibfnamefont
  {S.}~\bibnamefont {{Hu}}}, \ and\ \bibinfo {author} {\bibfnamefont
  {J.}~\bibnamefont {{Chen}}},\ }\href {\doibase 10.1103/PhysRevA.94.043419}
  {\bibfield  {journal} {\bibinfo  {journal} {\pra}\ }\textbf {\bibinfo
  {volume} {94}},\ \bibinfo {eid} {043419} (\bibinfo {year} {2016})},\ \Eprint
  {http://arxiv.org/abs/1608.04860} {arXiv:1608.04860 [physics.atom-ph]}
  \BibitemShut {NoStop}%
\bibitem [{\citenamefont {{Rudenko}}\ \emph {et~al.}(2004)\citenamefont
  {{Rudenko}}, \citenamefont {{Zrost}}, \citenamefont {{Schr{\"o}ter}},
  \citenamefont {{de Jesus}}, \citenamefont {{Feuerstein}}, \citenamefont
  {{Moshammer}},\ and\ \citenamefont {{Ullrich}}}]{Rudenko04}%
  \BibitemOpen
  \bibfield  {author} {\bibinfo {author} {\bibfnamefont {A.}~\bibnamefont
  {{Rudenko}}}, \bibinfo {author} {\bibfnamefont {K.}~\bibnamefont {{Zrost}}},
  \bibinfo {author} {\bibfnamefont {C.~D.}\ \bibnamefont {{Schr{\"o}ter}}},
  \bibinfo {author} {\bibfnamefont {V.~L.~B.}\ \bibnamefont {{de Jesus}}},
  \bibinfo {author} {\bibfnamefont {B.}~\bibnamefont {{Feuerstein}}}, \bibinfo
  {author} {\bibfnamefont {R.}~\bibnamefont {{Moshammer}}}, \ and\ \bibinfo
  {author} {\bibfnamefont {J.}~\bibnamefont {{Ullrich}}},\ }\href {\doibase
  10.1088/0953-4075/37/24/L03} {\bibfield  {journal} {\bibinfo  {journal}
  {Journal of Physics B Atomic Molecular Physics}\ }\textbf {\bibinfo {volume}
  {37}},\ \bibinfo {pages} {L407} (\bibinfo {year} {2004})},\ \Eprint
  {http://arxiv.org/abs/physics/0408064} {arXiv:physics/0408064
  [physics.atom-ph]} \BibitemShut {NoStop}%
\bibitem [{\citenamefont {{Maharjan}}\ \emph {et~al.}(2006)\citenamefont
  {{Maharjan}}, \citenamefont {{Alnaser}}, \citenamefont {{Litvinyuk}},
  \citenamefont {{Ranitovic}},\ and\ \citenamefont {{Cocke}}}]{Maharjan06}%
  \BibitemOpen
  \bibfield  {author} {\bibinfo {author} {\bibfnamefont {C.~M.}\ \bibnamefont
  {{Maharjan}}}, \bibinfo {author} {\bibfnamefont {A.~S.}\ \bibnamefont
  {{Alnaser}}}, \bibinfo {author} {\bibfnamefont {I.}~\bibnamefont
  {{Litvinyuk}}}, \bibinfo {author} {\bibfnamefont {P.}~\bibnamefont
  {{Ranitovic}}}, \ and\ \bibinfo {author} {\bibfnamefont {C.~L.}\ \bibnamefont
  {{Cocke}}},\ }\href {\doibase 10.1088/0953-4075/39/8/013} {\bibfield
  {journal} {\bibinfo  {journal} {Journal of Physics B Atomic Molecular
  Physics}\ }\textbf {\bibinfo {volume} {39}},\ \bibinfo {pages} {1955}
  (\bibinfo {year} {2006})}\BibitemShut {NoStop}%
\bibitem [{\citenamefont {Arb\'o}\ \emph
  {et~al.}(2006{\natexlab{b}})\citenamefont {Arb\'o}, \citenamefont {Yoshida},
  \citenamefont {Persson}, \citenamefont {Dimitriou},\ and\ \citenamefont
  {Burgd\"orfer}}]{Arbo06a}%
  \BibitemOpen
  \bibfield  {author} {\bibinfo {author} {\bibfnamefont {D.~G.}\ \bibnamefont
  {Arb\'o}}, \bibinfo {author} {\bibfnamefont {S.}~\bibnamefont {Yoshida}},
  \bibinfo {author} {\bibfnamefont {E.}~\bibnamefont {Persson}}, \bibinfo
  {author} {\bibfnamefont {K.~I.}\ \bibnamefont {Dimitriou}}, \ and\ \bibinfo
  {author} {\bibfnamefont {J.}~\bibnamefont {Burgd\"orfer}},\ }\href {\doibase
  10.1103/PhysRevLett.96.143003} {\bibfield  {journal} {\bibinfo  {journal}
  {Phys. Rev. Lett.}\ }\textbf {\bibinfo {volume} {96}},\ \bibinfo {pages}
  {143003} (\bibinfo {year} {2006}{\natexlab{b}})}\BibitemShut {NoStop}%
\bibitem [{\citenamefont {Arb\'o}\ \emph {et~al.}(2008)\citenamefont {Arb\'o},
  \citenamefont {Dimitriou}, \citenamefont {Persson},\ and\ \citenamefont
  {Burgd\"orfer}}]{Arbo08b}%
  \BibitemOpen
  \bibfield  {author} {\bibinfo {author} {\bibfnamefont {D.~G.}\ \bibnamefont
  {Arb\'o}}, \bibinfo {author} {\bibfnamefont {K.~I.}\ \bibnamefont
  {Dimitriou}}, \bibinfo {author} {\bibfnamefont {E.}~\bibnamefont {Persson}},
  \ and\ \bibinfo {author} {\bibfnamefont {J.}~\bibnamefont {Burgd\"orfer}},\
  }\href {\doibase 10.1103/PhysRevA.78.013406} {\bibfield  {journal} {\bibinfo
  {journal} {Phys. Rev. A}\ }\textbf {\bibinfo {volume} {78}},\ \bibinfo
  {pages} {013406} (\bibinfo {year} {2008})}\BibitemShut {NoStop}%
\bibitem [{\citenamefont {{Borb{\'e}ly}}\ \emph {et~al.}(2013)\citenamefont
  {{Borb{\'e}ly}}, \citenamefont {{T{\'o}th}}, \citenamefont {{T{\H
  o}k{\'e}si}},\ and\ \citenamefont {{Nagy}}}]{Borbely13}%
  \BibitemOpen
  \bibfield  {author} {\bibinfo {author} {\bibfnamefont {S.}~\bibnamefont
  {{Borb{\'e}ly}}}, \bibinfo {author} {\bibfnamefont {A.}~\bibnamefont
  {{T{\'o}th}}}, \bibinfo {author} {\bibfnamefont {K.}~\bibnamefont {{T{\H
  o}k{\'e}si}}}, \ and\ \bibinfo {author} {\bibfnamefont {L.}~\bibnamefont
  {{Nagy}}},\ }\href {\doibase 10.1103/PhysRevA.87.013405} {\bibfield
  {journal} {\bibinfo  {journal} {Phys. Rev. A}\ }\textbf {\bibinfo {volume}
  {87}},\ \bibinfo {eid} {013405} (\bibinfo {year} {2013})}\BibitemShut
  {NoStop}%
\bibitem [{\citenamefont {{Yan}}\ \emph {et~al.}(2010)\citenamefont {{Yan}},
  \citenamefont {{Popruzhenko}}, \citenamefont {{Vrakking}},\ and\
  \citenamefont {{Bauer}}}]{Yan10}%
  \BibitemOpen
  \bibfield  {author} {\bibinfo {author} {\bibfnamefont {T.-M.}\ \bibnamefont
  {{Yan}}}, \bibinfo {author} {\bibfnamefont {S.~V.}\ \bibnamefont
  {{Popruzhenko}}}, \bibinfo {author} {\bibfnamefont {M.~J.~J.}\ \bibnamefont
  {{Vrakking}}}, \ and\ \bibinfo {author} {\bibfnamefont {D.}~\bibnamefont
  {{Bauer}}},\ }\href {\doibase 10.1103/PhysRevLett.105.253002} {\bibfield
  {journal} {\bibinfo  {journal} {\prl}\ }\textbf {\bibinfo {volume} {105}},\
  \bibinfo {eid} {253002} (\bibinfo {year} {2010})},\ \Eprint
  {http://arxiv.org/abs/1008.3144} {arXiv:1008.3144 [physics.atom-ph]}
  \BibitemShut {NoStop}%
\bibitem [{\citenamefont {{Huismans}}\ \emph {et~al.}(2011)\citenamefont
  {{Huismans}}, \citenamefont {{Rouz{\'e}e}}, \citenamefont {{Gijsbertsen}},
  \citenamefont {{Jungmann}}, \citenamefont {{Smolkowska}}, \citenamefont
  {{Logman}}, \citenamefont {{L{\'e}pine}}, \citenamefont {{Cauchy}},
  \citenamefont {{Zamith}}, \citenamefont {{Marchenko}}, \citenamefont
  {{Bakker}}, \citenamefont {{Berden}}, \citenamefont {{Redlich}},
  \citenamefont {{van der Meer}}, \citenamefont {{Muller}}, \citenamefont
  {{Vermin}}, \citenamefont {{Schafer}}, \citenamefont {{Spanner}},
  \citenamefont {{Ivanov}}, \citenamefont {{Smirnova}}, \citenamefont
  {{Bauer}}, \citenamefont {{Popruzhenko}},\ and\ \citenamefont
  {{Vrakking}}}]{Huismans11}%
  \BibitemOpen
  \bibfield  {author} {\bibinfo {author} {\bibfnamefont {Y.}~\bibnamefont
  {{Huismans}}}, \bibinfo {author} {\bibfnamefont {A.}~\bibnamefont
  {{Rouz{\'e}e}}}, \bibinfo {author} {\bibfnamefont {A.}~\bibnamefont
  {{Gijsbertsen}}}, \bibinfo {author} {\bibfnamefont {J.~H.}\ \bibnamefont
  {{Jungmann}}}, \bibinfo {author} {\bibfnamefont {A.~S.}\ \bibnamefont
  {{Smolkowska}}}, \bibinfo {author} {\bibfnamefont {P.~S.~W.~M.}\ \bibnamefont
  {{Logman}}}, \bibinfo {author} {\bibfnamefont {F.}~\bibnamefont
  {{L{\'e}pine}}}, \bibinfo {author} {\bibfnamefont {C.}~\bibnamefont
  {{Cauchy}}}, \bibinfo {author} {\bibfnamefont {S.}~\bibnamefont {{Zamith}}},
  \bibinfo {author} {\bibfnamefont {T.}~\bibnamefont {{Marchenko}}}, \bibinfo
  {author} {\bibfnamefont {J.~M.}\ \bibnamefont {{Bakker}}}, \bibinfo {author}
  {\bibfnamefont {G.}~\bibnamefont {{Berden}}}, \bibinfo {author}
  {\bibfnamefont {B.}~\bibnamefont {{Redlich}}}, \bibinfo {author}
  {\bibfnamefont {A.~F.~G.}\ \bibnamefont {{van der Meer}}}, \bibinfo {author}
  {\bibfnamefont {H.~G.}\ \bibnamefont {{Muller}}}, \bibinfo {author}
  {\bibfnamefont {W.}~\bibnamefont {{Vermin}}}, \bibinfo {author}
  {\bibfnamefont {K.~J.}\ \bibnamefont {{Schafer}}}, \bibinfo {author}
  {\bibfnamefont {M.}~\bibnamefont {{Spanner}}}, \bibinfo {author}
  {\bibfnamefont {M.~Y.}\ \bibnamefont {{Ivanov}}}, \bibinfo {author}
  {\bibfnamefont {O.}~\bibnamefont {{Smirnova}}}, \bibinfo {author}
  {\bibfnamefont {D.}~\bibnamefont {{Bauer}}}, \bibinfo {author} {\bibfnamefont
  {S.~V.}\ \bibnamefont {{Popruzhenko}}}, \ and\ \bibinfo {author}
  {\bibfnamefont {M.~J.~J.}\ \bibnamefont {{Vrakking}}},\ }\href {\doibase
  10.1126/science.1198450} {\bibfield  {journal} {\bibinfo  {journal}
  {Science}\ }\textbf {\bibinfo {volume} {331}},\ \bibinfo {pages} {61}
  (\bibinfo {year} {2011})}\BibitemShut {NoStop}%
\bibitem [{\citenamefont {{Huismans}}\ \emph {et~al.}(2012)\citenamefont
  {{Huismans}}, \citenamefont {{Gijsbertsen}}, \citenamefont {{Smolkowska}},
  \citenamefont {{Jungmann}}, \citenamefont {{Rouz{\'e}e}}, \citenamefont
  {{Logman}}, \citenamefont {{L{\'e}pine}}, \citenamefont {{Cauchy}},
  \citenamefont {{Zamith}}, \citenamefont {{Marchenko}}, \citenamefont
  {{Bakker}}, \citenamefont {{Berden}}, \citenamefont {{Redlich}},
  \citenamefont {{van der Meer}}, \citenamefont {{Ivanov}}, \citenamefont
  {{Yan}}, \citenamefont {{Bauer}}, \citenamefont {{Smirnova}},\ and\
  \citenamefont {{Vrakking}}}]{Huismans12}%
  \BibitemOpen
  \bibfield  {author} {\bibinfo {author} {\bibfnamefont {Y.}~\bibnamefont
  {{Huismans}}}, \bibinfo {author} {\bibfnamefont {A.}~\bibnamefont
  {{Gijsbertsen}}}, \bibinfo {author} {\bibfnamefont {A.~S.}\ \bibnamefont
  {{Smolkowska}}}, \bibinfo {author} {\bibfnamefont {J.~H.}\ \bibnamefont
  {{Jungmann}}}, \bibinfo {author} {\bibfnamefont {A.}~\bibnamefont
  {{Rouz{\'e}e}}}, \bibinfo {author} {\bibfnamefont {P.~S.~W.~M.}\ \bibnamefont
  {{Logman}}}, \bibinfo {author} {\bibfnamefont {F.}~\bibnamefont
  {{L{\'e}pine}}}, \bibinfo {author} {\bibfnamefont {C.}~\bibnamefont
  {{Cauchy}}}, \bibinfo {author} {\bibfnamefont {S.}~\bibnamefont {{Zamith}}},
  \bibinfo {author} {\bibfnamefont {T.}~\bibnamefont {{Marchenko}}}, \bibinfo
  {author} {\bibfnamefont {J.~M.}\ \bibnamefont {{Bakker}}}, \bibinfo {author}
  {\bibfnamefont {G.}~\bibnamefont {{Berden}}}, \bibinfo {author}
  {\bibfnamefont {B.}~\bibnamefont {{Redlich}}}, \bibinfo {author}
  {\bibfnamefont {A.~F.~G.}\ \bibnamefont {{van der Meer}}}, \bibinfo {author}
  {\bibfnamefont {M.~Y.}\ \bibnamefont {{Ivanov}}}, \bibinfo {author}
  {\bibfnamefont {T.-M.}\ \bibnamefont {{Yan}}}, \bibinfo {author}
  {\bibfnamefont {D.}~\bibnamefont {{Bauer}}}, \bibinfo {author} {\bibfnamefont
  {O.}~\bibnamefont {{Smirnova}}}, \ and\ \bibinfo {author} {\bibfnamefont
  {M.~J.~J.}\ \bibnamefont {{Vrakking}}},\ }\href {\doibase
  10.1103/PhysRevLett.109.013002} {\bibfield  {journal} {\bibinfo  {journal}
  {Physical Review Letters}\ }\textbf {\bibinfo {volume} {109}},\ \bibinfo
  {eid} {013002} (\bibinfo {year} {2012})}\BibitemShut {NoStop}%
\bibitem [{\citenamefont {{Song}}\ \emph {et~al.}(2016)\citenamefont {{Song}},
  \citenamefont {{Lin}}, \citenamefont {{Sheng}}, \citenamefont {{Liu}},
  \citenamefont {{Chen}}, \citenamefont {{Yang}}, \citenamefont {{Hu}},
  \citenamefont {{Lin}},\ and\ \citenamefont {{Chen}}}]{Song16}%
  \BibitemOpen
  \bibfield  {author} {\bibinfo {author} {\bibfnamefont {X.}~\bibnamefont
  {{Song}}}, \bibinfo {author} {\bibfnamefont {C.}~\bibnamefont {{Lin}}},
  \bibinfo {author} {\bibfnamefont {Z.}~\bibnamefont {{Sheng}}}, \bibinfo
  {author} {\bibfnamefont {P.}~\bibnamefont {{Liu}}}, \bibinfo {author}
  {\bibfnamefont {Z.}~\bibnamefont {{Chen}}}, \bibinfo {author} {\bibfnamefont
  {W.}~\bibnamefont {{Yang}}}, \bibinfo {author} {\bibfnamefont
  {S.}~\bibnamefont {{Hu}}}, \bibinfo {author} {\bibfnamefont {C.~D.}\
  \bibnamefont {{Lin}}}, \ and\ \bibinfo {author} {\bibfnamefont
  {J.}~\bibnamefont {{Chen}}},\ }\href {\doibase 10.1038/srep28392} {\bibfield
  {journal} {\bibinfo  {journal} {Scientific Reports}\ }\textbf {\bibinfo
  {volume} {6}},\ \bibinfo {eid} {28392} (\bibinfo {year} {2016})},\ \Eprint
  {http://arxiv.org/abs/1602.06019} {arXiv:1602.06019 [physics.atom-ph]}
  \BibitemShut {NoStop}%
\bibitem [{\citenamefont {{Lai}}\ \emph {et~al.}(2017)\citenamefont {{Lai}},
  \citenamefont {{Yu}}, \citenamefont {{Huang}}, \citenamefont {{Hua}},
  \citenamefont {{Gong}}, \citenamefont {{Quan}}, \citenamefont {{Faria}},\
  and\ \citenamefont {{Liu}}}]{Lai17}%
  \BibitemOpen
  \bibfield  {author} {\bibinfo {author} {\bibfnamefont {X.}~\bibnamefont
  {{Lai}}}, \bibinfo {author} {\bibfnamefont {S.}~\bibnamefont {{Yu}}},
  \bibinfo {author} {\bibfnamefont {Y.}~\bibnamefont {{Huang}}}, \bibinfo
  {author} {\bibfnamefont {L.}~\bibnamefont {{Hua}}}, \bibinfo {author}
  {\bibfnamefont {C.}~\bibnamefont {{Gong}}}, \bibinfo {author} {\bibfnamefont
  {W.}~\bibnamefont {{Quan}}}, \bibinfo {author} {\bibfnamefont {C.~F. d.~M.}\
  \bibnamefont {{Faria}}}, \ and\ \bibinfo {author} {\bibfnamefont
  {X.}~\bibnamefont {{Liu}}},\ }\href {\doibase 10.1103/PhysRevA.96.013414}
  {\bibfield  {journal} {\bibinfo  {journal} {\pra}\ }\textbf {\bibinfo
  {volume} {96}},\ \bibinfo {eid} {013414} (\bibinfo {year} {2017})},\ \Eprint
  {http://arxiv.org/abs/1703.04123} {arXiv:1703.04123 [physics.atom-ph]}
  \BibitemShut {NoStop}%
\bibitem [{\citenamefont {Shvetsov-Shilovski}\ and\ \citenamefont
  {Lein}(2018)}]{Shilovski18}%
  \BibitemOpen
  \bibfield  {author} {\bibinfo {author} {\bibfnamefont {N.~I.}\ \bibnamefont
  {Shvetsov-Shilovski}}\ and\ \bibinfo {author} {\bibfnamefont
  {M.}~\bibnamefont {Lein}},\ }\href {\doibase 10.1103/PhysRevA.97.013411}
  {\bibfield  {journal} {\bibinfo  {journal} {Phys. Rev. A}\ }\textbf {\bibinfo
  {volume} {97}},\ \bibinfo {pages} {013411} (\bibinfo {year}
  {2018})}\BibitemShut {NoStop}%
\bibitem [{\citenamefont {{Tan}}\ \emph {et~al.}(2019)\citenamefont {{Tan}},
  \citenamefont {{Zhou}}, \citenamefont {{He}}, \citenamefont {{Ke}},
  \citenamefont {{Liang}}, \citenamefont {{Li}}, \citenamefont {{Li}},\ and\
  \citenamefont {{Lu}}}]{Tan19}%
  \BibitemOpen
  \bibfield  {author} {\bibinfo {author} {\bibfnamefont {J.}~\bibnamefont
  {{Tan}}}, \bibinfo {author} {\bibfnamefont {Y.}~\bibnamefont {{Zhou}}},
  \bibinfo {author} {\bibfnamefont {M.}~\bibnamefont {{He}}}, \bibinfo {author}
  {\bibfnamefont {Q.}~\bibnamefont {{Ke}}}, \bibinfo {author} {\bibfnamefont
  {J.}~\bibnamefont {{Liang}}}, \bibinfo {author} {\bibfnamefont
  {Y.}~\bibnamefont {{Li}}}, \bibinfo {author} {\bibfnamefont {M.}~\bibnamefont
  {{Li}}}, \ and\ \bibinfo {author} {\bibfnamefont {P.}~\bibnamefont {{Lu}}},\
  }\href {\doibase 10.1103/PhysRevA.99.033402} {\bibfield  {journal} {\bibinfo
  {journal} {\pra}\ }\textbf {\bibinfo {volume} {99}},\ \bibinfo {eid} {033402}
  (\bibinfo {year} {2019})}\BibitemShut {NoStop}%
\bibitem [{\citenamefont {Porat}\ \emph {et~al.}(2018)\citenamefont {Porat},
  \citenamefont {Alon}, \citenamefont {Rozen}, \citenamefont {Pedatzur},
  \citenamefont {Kr{\"u}ger}, \citenamefont {Azoury}, \citenamefont {Natan},
  \citenamefont {Orenstein}, \citenamefont {Bruner}, \citenamefont {Vrakking}
  \emph {et~al.}}]{Porat18}%
  \BibitemOpen
  \bibfield  {author} {\bibinfo {author} {\bibfnamefont {G.}~\bibnamefont
  {Porat}}, \bibinfo {author} {\bibfnamefont {G.}~\bibnamefont {Alon}},
  \bibinfo {author} {\bibfnamefont {S.}~\bibnamefont {Rozen}}, \bibinfo
  {author} {\bibfnamefont {O.}~\bibnamefont {Pedatzur}}, \bibinfo {author}
  {\bibfnamefont {M.}~\bibnamefont {Kr{\"u}ger}}, \bibinfo {author}
  {\bibfnamefont {D.}~\bibnamefont {Azoury}}, \bibinfo {author} {\bibfnamefont
  {A.}~\bibnamefont {Natan}}, \bibinfo {author} {\bibfnamefont
  {G.}~\bibnamefont {Orenstein}}, \bibinfo {author} {\bibfnamefont
  {B.}~\bibnamefont {Bruner}}, \bibinfo {author} {\bibfnamefont
  {M.}~\bibnamefont {Vrakking}},  \emph {et~al.},\ }\href@noop {} {\bibfield
  {journal} {\bibinfo  {journal} {Nature communications}\ }\textbf {\bibinfo
  {volume} {9}},\ \bibinfo {pages} {2805} (\bibinfo {year} {2018})}\BibitemShut
  {NoStop}%
\bibitem [{\citenamefont {{Raz}}\ \emph {et~al.}(2012)\citenamefont {{Raz}},
  \citenamefont {{Pedatzur}}, \citenamefont {{Bruner}},\ and\ \citenamefont
  {{Dudovich}}}]{Raz12}%
  \BibitemOpen
  \bibfield  {author} {\bibinfo {author} {\bibfnamefont {O.}~\bibnamefont
  {{Raz}}}, \bibinfo {author} {\bibfnamefont {O.}~\bibnamefont {{Pedatzur}}},
  \bibinfo {author} {\bibfnamefont {B.~D.}\ \bibnamefont {{Bruner}}}, \ and\
  \bibinfo {author} {\bibfnamefont {N.}~\bibnamefont {{Dudovich}}},\ }\href
  {\doibase 10.1038/nphoton.2011.353} {\bibfield  {journal} {\bibinfo
  {journal} {Nature Photonics}\ }\textbf {\bibinfo {volume} {6}},\ \bibinfo
  {pages} {170} (\bibinfo {year} {2012})}\BibitemShut {NoStop}%
\bibitem [{\citenamefont {{Xia}}\ \emph {et~al.}(2018)\citenamefont {{Xia}},
  \citenamefont {{Tao}}, \citenamefont {{Cai}}, \citenamefont {{Fu}},\ and\
  \citenamefont {{Liu}}}]{Xia18}%
  \BibitemOpen
  \bibfield  {author} {\bibinfo {author} {\bibfnamefont {Q.~Z.}\ \bibnamefont
  {{Xia}}}, \bibinfo {author} {\bibfnamefont {J.~F.}\ \bibnamefont {{Tao}}},
  \bibinfo {author} {\bibfnamefont {J.}~\bibnamefont {{Cai}}}, \bibinfo
  {author} {\bibfnamefont {L.~B.}\ \bibnamefont {{Fu}}}, \ and\ \bibinfo
  {author} {\bibfnamefont {J.}~\bibnamefont {{Liu}}},\ }\href {\doibase
  10.1103/PhysRevLett.121.143201} {\bibfield  {journal} {\bibinfo  {journal}
  {Physical Review Letters}\ }\textbf {\bibinfo {volume} {121}},\ \bibinfo
  {eid} {143201} (\bibinfo {year} {2018})},\ \Eprint
  {http://arxiv.org/abs/1708.04374} {arXiv:1708.04374 [physics.atom-ph]}
  \BibitemShut {NoStop}%
\bibitem [{\citenamefont {{Xie}}\ \emph {et~al.}(2016)\citenamefont {{Xie}},
  \citenamefont {{Li}}, \citenamefont {{Li}}, \citenamefont {{Zhou}},\ and\
  \citenamefont {{Lu}}}]{Xie16}%
  \BibitemOpen
  \bibfield  {author} {\bibinfo {author} {\bibfnamefont {H.}~\bibnamefont
  {{Xie}}}, \bibinfo {author} {\bibfnamefont {M.}~\bibnamefont {{Li}}},
  \bibinfo {author} {\bibfnamefont {Y.}~\bibnamefont {{Li}}}, \bibinfo {author}
  {\bibfnamefont {Y.}~\bibnamefont {{Zhou}}}, \ and\ \bibinfo {author}
  {\bibfnamefont {P.}~\bibnamefont {{Lu}}},\ }\href {\doibase
  10.1364/OE.24.027726} {\bibfield  {journal} {\bibinfo  {journal} {Optics
  Express}\ }\textbf {\bibinfo {volume} {24}},\ \bibinfo {pages} {27726}
  (\bibinfo {year} {2016})}\BibitemShut {NoStop}%
\bibitem [{\citenamefont {{Shvetsov-Shilovski}}\ \emph
  {et~al.}(2016)\citenamefont {{Shvetsov-Shilovski}}, \citenamefont {{Lein}},
  \citenamefont {{Madsen}}, \citenamefont {{R{\"a}s{\"a}nen}}, \citenamefont
  {{Lemell}}, \citenamefont {{Burgd{\"o}rfer}}, \citenamefont {{Arb{\'o}}},\
  and\ \citenamefont {{T{\H o}k{\'e}si}}}]{Shilovski16}%
  \BibitemOpen
  \bibfield  {author} {\bibinfo {author} {\bibfnamefont {N.~I.}\ \bibnamefont
  {{Shvetsov-Shilovski}}}, \bibinfo {author} {\bibfnamefont {M.}~\bibnamefont
  {{Lein}}}, \bibinfo {author} {\bibfnamefont {L.~B.}\ \bibnamefont
  {{Madsen}}}, \bibinfo {author} {\bibfnamefont {E.}~\bibnamefont
  {{R{\"a}s{\"a}nen}}}, \bibinfo {author} {\bibfnamefont {C.}~\bibnamefont
  {{Lemell}}}, \bibinfo {author} {\bibfnamefont {J.}~\bibnamefont
  {{Burgd{\"o}rfer}}}, \bibinfo {author} {\bibfnamefont {D.~G.}\ \bibnamefont
  {{Arb{\'o}}}}, \ and\ \bibinfo {author} {\bibfnamefont {K.}~\bibnamefont
  {{T{\H o}k{\'e}si}}},\ }\href {\doibase 10.1103/PhysRevA.94.013415}
  {\bibfield  {journal} {\bibinfo  {journal} {Phys. Rev. A}\ }\textbf {\bibinfo
  {volume} {94}},\ \bibinfo {eid} {013415} (\bibinfo {year} {2016})},\ \Eprint
  {http://arxiv.org/abs/1604.05123} {arXiv:1604.05123 [physics.atom-ph]}
  \BibitemShut {NoStop}%
\bibitem [{\citenamefont {{Song}}\ \emph {et~al.}(2017)\citenamefont {{Song}},
  \citenamefont {{Xu}}, \citenamefont {{Lin}}, \citenamefont {{Sheng}},
  \citenamefont {{Liu}}, \citenamefont {{Yu}}, \citenamefont {{Zhang}},
  \citenamefont {{Yang}}, \citenamefont {{Hu}}, \citenamefont {{Chen}},
  \citenamefont {{Xu}}, \citenamefont {{Chen}}, \citenamefont {{Quan}},\ and\
  \citenamefont {{Liu}}}]{Song17}%
  \BibitemOpen
  \bibfield  {author} {\bibinfo {author} {\bibfnamefont {X.}~\bibnamefont
  {{Song}}}, \bibinfo {author} {\bibfnamefont {J.}~\bibnamefont {{Xu}}},
  \bibinfo {author} {\bibfnamefont {C.}~\bibnamefont {{Lin}}}, \bibinfo
  {author} {\bibfnamefont {Z.}~\bibnamefont {{Sheng}}}, \bibinfo {author}
  {\bibfnamefont {P.}~\bibnamefont {{Liu}}}, \bibinfo {author} {\bibfnamefont
  {X.}~\bibnamefont {{Yu}}}, \bibinfo {author} {\bibfnamefont {H.}~\bibnamefont
  {{Zhang}}}, \bibinfo {author} {\bibfnamefont {W.}~\bibnamefont {{Yang}}},
  \bibinfo {author} {\bibfnamefont {S.}~\bibnamefont {{Hu}}}, \bibinfo {author}
  {\bibfnamefont {J.}~\bibnamefont {{Chen}}}, \bibinfo {author} {\bibfnamefont
  {S.}~\bibnamefont {{Xu}}}, \bibinfo {author} {\bibfnamefont {Y.}~\bibnamefont
  {{Chen}}}, \bibinfo {author} {\bibfnamefont {W.}~\bibnamefont {{Quan}}}, \
  and\ \bibinfo {author} {\bibfnamefont {X.}~\bibnamefont {{Liu}}},\ }\href
  {\doibase 10.1103/PhysRevA.95.033426} {\bibfield  {journal} {\bibinfo
  {journal} {\pra}\ }\textbf {\bibinfo {volume} {95}},\ \bibinfo {eid} {033426}
  (\bibinfo {year} {2017})},\ \Eprint {http://arxiv.org/abs/1602.05668}
  {arXiv:1602.05668 [physics.atom-ph]} \BibitemShut {NoStop}%
\bibitem [{\citenamefont {{Ni}}\ \emph {et~al.}(2018)\citenamefont {{Ni}},
  \citenamefont {{Eicke}}, \citenamefont {{Ruiz}}, \citenamefont {{Cai}},
  \citenamefont {{Oppermann}}, \citenamefont {{Shvetsov-Shilovski}},\ and\
  \citenamefont {{Pi}}}]{Ni18}%
  \BibitemOpen
  \bibfield  {author} {\bibinfo {author} {\bibfnamefont {H.}~\bibnamefont
  {{Ni}}}, \bibinfo {author} {\bibfnamefont {N.}~\bibnamefont {{Eicke}}},
  \bibinfo {author} {\bibfnamefont {C.}~\bibnamefont {{Ruiz}}}, \bibinfo
  {author} {\bibfnamefont {J.}~\bibnamefont {{Cai}}}, \bibinfo {author}
  {\bibfnamefont {F.}~\bibnamefont {{Oppermann}}}, \bibinfo {author}
  {\bibfnamefont {N.~I.}\ \bibnamefont {{Shvetsov-Shilovski}}}, \ and\ \bibinfo
  {author} {\bibfnamefont {L.-W.}\ \bibnamefont {{Pi}}},\ }\href {\doibase
  10.1103/PhysRevA.98.013411} {\bibfield  {journal} {\bibinfo  {journal}
  {\pra}\ }\textbf {\bibinfo {volume} {98}},\ \bibinfo {eid} {013411} (\bibinfo
  {year} {2018})}\BibitemShut {NoStop}%
\bibitem [{\citenamefont {{Xiao}}\ \emph {et~al.}(2016)\citenamefont {{Xiao}},
  \citenamefont {{Wang}}, \citenamefont {{Xiong}},\ and\ \citenamefont
  {{Peng}}}]{Xiao16}%
  \BibitemOpen
  \bibfield  {author} {\bibinfo {author} {\bibfnamefont {X.-R.}\ \bibnamefont
  {{Xiao}}}, \bibinfo {author} {\bibfnamefont {M.-X.}\ \bibnamefont {{Wang}}},
  \bibinfo {author} {\bibfnamefont {W.-H.}\ \bibnamefont {{Xiong}}}, \ and\
  \bibinfo {author} {\bibfnamefont {L.-Y.}\ \bibnamefont {{Peng}}},\ }\href
  {\doibase 10.1103/PhysRevE.94.053310} {\bibfield  {journal} {\bibinfo
  {journal} {\pre}\ }\textbf {\bibinfo {volume} {94}},\ \bibinfo {eid} {053310}
  (\bibinfo {year} {2016})}\BibitemShut {NoStop}%
\bibitem [{\citenamefont {{Zagoya}}\ \emph {et~al.}(2014)\citenamefont
  {{Zagoya}}, \citenamefont {{Wu}}, \citenamefont {{Ronto}}, \citenamefont
  {{Shalashilin}},\ and\ \citenamefont {{Figueira de Morisson
  Faria}}}]{Zagoya14}%
  \BibitemOpen
  \bibfield  {author} {\bibinfo {author} {\bibfnamefont {C.}~\bibnamefont
  {{Zagoya}}}, \bibinfo {author} {\bibfnamefont {J.}~\bibnamefont {{Wu}}},
  \bibinfo {author} {\bibfnamefont {M.}~\bibnamefont {{Ronto}}}, \bibinfo
  {author} {\bibfnamefont {D.~V.}\ \bibnamefont {{Shalashilin}}}, \ and\
  \bibinfo {author} {\bibfnamefont {C.}~\bibnamefont {{Figueira de Morisson
  Faria}}},\ }\href {\doibase 10.1088/1367-2630/16/10/103040} {\bibfield
  {journal} {\bibinfo  {journal} {New Journal of Physics}\ }\textbf {\bibinfo
  {volume} {16}},\ \bibinfo {eid} {103040} (\bibinfo {year} {2014})},\ \Eprint
  {http://arxiv.org/abs/1405.2873} {arXiv:1405.2873 [physics.atom-ph]}
  \BibitemShut {NoStop}%
\bibitem [{\citenamefont {{Maxwell}}\ and\ \citenamefont {{Figueira de Morisson
  Faria}}(2018)}]{Maxwell18a}%
  \BibitemOpen
  \bibfield  {author} {\bibinfo {author} {\bibfnamefont {A.~S.}\ \bibnamefont
  {{Maxwell}}}\ and\ \bibinfo {author} {\bibfnamefont {C.}~\bibnamefont
  {{Figueira de Morisson Faria}}},\ }\href {\doibase 10.1088/1361-6455/aac164}
  {\bibfield  {journal} {\bibinfo  {journal} {Journal of Physics B Atomic
  Molecular Physics}\ }\textbf {\bibinfo {volume} {51}},\ \bibinfo {eid}
  {124001} (\bibinfo {year} {2018})},\ \Eprint
  {http://arxiv.org/abs/1802.00789} {arXiv:1802.00789 [physics.atom-ph]}
  \BibitemShut {NoStop}%
\bibitem [{\citenamefont {{Maxwell}}\ \emph {et~al.}(2018)\citenamefont
  {{Maxwell}}, \citenamefont {{Al-Jawahiry}}, \citenamefont {{Lai}},\ and\
  \citenamefont {{Figueira de Morisson Faria}}}]{Maxwell18b}%
  \BibitemOpen
  \bibfield  {author} {\bibinfo {author} {\bibfnamefont {A.~S.}\ \bibnamefont
  {{Maxwell}}}, \bibinfo {author} {\bibfnamefont {A.}~\bibnamefont
  {{Al-Jawahiry}}}, \bibinfo {author} {\bibfnamefont {X.~Y.}\ \bibnamefont
  {{Lai}}}, \ and\ \bibinfo {author} {\bibfnamefont {C.}~\bibnamefont
  {{Figueira de Morisson Faria}}},\ }\href {\doibase 10.1088/1361-6455/aa9e81}
  {\bibfield  {journal} {\bibinfo  {journal} {Journal of Physics B Atomic
  Molecular Physics}\ }\textbf {\bibinfo {volume} {51}},\ \bibinfo {eid}
  {044004} (\bibinfo {year} {2018})},\ \Eprint
  {http://arxiv.org/abs/1709.05973} {arXiv:1709.05973 [physics.atom-ph]}
  \BibitemShut {NoStop}%
\bibitem [{\citenamefont {Tong}\ and\ \citenamefont {Chu}(1997)}]{Tong97}%
  \BibitemOpen
  \bibfield  {author} {\bibinfo {author} {\bibfnamefont {X.~M.}\ \bibnamefont
  {Tong}}\ and\ \bibinfo {author} {\bibfnamefont {S.~I.}\ \bibnamefont {Chu}},\
  }\href {http://dx.doi.org/10.1016/S0301-0104(97)00063-3} {\bibfield
  {journal} {\bibinfo  {journal} {Chem. Phys.}\ }\textbf {\bibinfo {volume}
  {217}},\ \bibinfo {pages} {119} (\bibinfo {year} {1997})}\BibitemShut
  {NoStop}%
\bibitem [{\citenamefont {Tong}\ and\ \citenamefont {Chu}(2000)}]{Tong00}%
  \BibitemOpen
  \bibfield  {author} {\bibinfo {author} {\bibfnamefont {X.-M.}\ \bibnamefont
  {Tong}}\ and\ \bibinfo {author} {\bibfnamefont {S.-I.}\ \bibnamefont {Chu}},\
  }\href {\doibase 10.1103/PhysRevA.61.031401} {\bibfield  {journal} {\bibinfo
  {journal} {Phys. Rev. A}\ }\textbf {\bibinfo {volume} {61}},\ \bibinfo
  {pages} {031401} (\bibinfo {year} {2000})}\BibitemShut {NoStop}%
\bibitem [{\citenamefont {Tong}\ and\ \citenamefont {Lin}(2005)}]{Tong05}%
  \BibitemOpen
  \bibfield  {author} {\bibinfo {author} {\bibfnamefont {X.~M.}\ \bibnamefont
  {Tong}}\ and\ \bibinfo {author} {\bibfnamefont {C.~D.}\ \bibnamefont {Lin}},\
  }\href {http://stacks.iop.org/0953-4075/38/i=15/a=001} {\bibfield  {journal}
  {\bibinfo  {journal} {Journal of Physics B: Atomic, Molecular and Optical
  Physics}\ }\textbf {\bibinfo {volume} {38}},\ \bibinfo {pages} {2593}
  (\bibinfo {year} {2005})}\BibitemShut {NoStop}%
\bibitem [{\citenamefont {{Shvetsov-Shilovski}}\ \emph
  {et~al.}(2012)\citenamefont {{Shvetsov-Shilovski}}, \citenamefont
  {{Dimitrovski}},\ and\ \citenamefont {{Madsen}}}]{Shilovski12}%
  \BibitemOpen
  \bibfield  {author} {\bibinfo {author} {\bibfnamefont {N.~I.}\ \bibnamefont
  {{Shvetsov-Shilovski}}}, \bibinfo {author} {\bibfnamefont {D.}~\bibnamefont
  {{Dimitrovski}}}, \ and\ \bibinfo {author} {\bibfnamefont {L.~B.}\
  \bibnamefont {{Madsen}}},\ }\href {\doibase 10.1103/PhysRevA.85.023428}
  {\bibfield  {journal} {\bibinfo  {journal} {Phys. Rev. A}\ }\textbf {\bibinfo
  {volume} {85}},\ \bibinfo {eid} {023428} (\bibinfo {year}
  {2012})}\BibitemShut {NoStop}%
\bibitem [{\citenamefont {Li}\ \emph {et~al.}(2014)\citenamefont {Li},
  \citenamefont {Geng}, \citenamefont {Liu}, \citenamefont {Deng},
  \citenamefont {Wu}, \citenamefont {Peng}, \citenamefont {Gong},\ and\
  \citenamefont {Liu}}]{Li14}%
  \BibitemOpen
  \bibfield  {author} {\bibinfo {author} {\bibfnamefont {M.}~\bibnamefont
  {Li}}, \bibinfo {author} {\bibfnamefont {J.-W.}\ \bibnamefont {Geng}},
  \bibinfo {author} {\bibfnamefont {H.}~\bibnamefont {Liu}}, \bibinfo {author}
  {\bibfnamefont {Y.}~\bibnamefont {Deng}}, \bibinfo {author} {\bibfnamefont
  {C.}~\bibnamefont {Wu}}, \bibinfo {author} {\bibfnamefont {L.-Y.}\
  \bibnamefont {Peng}}, \bibinfo {author} {\bibfnamefont {Q.}~\bibnamefont
  {Gong}}, \ and\ \bibinfo {author} {\bibfnamefont {Y.}~\bibnamefont {Liu}},\
  }\href {\doibase 10.1103/PhysRevLett.112.113002} {\bibfield  {journal}
  {\bibinfo  {journal} {Phys. Rev. Lett.}\ }\textbf {\bibinfo {volume} {112}},\
  \bibinfo {pages} {113002} (\bibinfo {year} {2014})}\BibitemShut {NoStop}%
\bibitem [{\citenamefont {{Dewangan}}\ and\ \citenamefont
  {{Eichler}}(1994)}]{Dewangan94}%
  \BibitemOpen
  \bibfield  {author} {\bibinfo {author} {\bibfnamefont {D.~P.}\ \bibnamefont
  {{Dewangan}}}\ and\ \bibinfo {author} {\bibfnamefont {J.}~\bibnamefont
  {{Eichler}}},\ }\href {\doibase 10.1016/0370-1573(94)90012-4} {\bibfield
  {journal} {\bibinfo  {journal} {Physics Reports}\ }\textbf {\bibinfo {volume}
  {247}},\ \bibinfo {pages} {59} (\bibinfo {year} {1994})}\BibitemShut
  {NoStop}%
\bibitem [{\citenamefont {{Macri}}\ \emph {et~al.}(2003)\citenamefont
  {{Macri}}, \citenamefont {{Miraglia}},\ and\ \citenamefont
  {{Gravielle}}}]{Macri03}%
  \BibitemOpen
  \bibfield  {author} {\bibinfo {author} {\bibfnamefont {P.~A.}\ \bibnamefont
  {{Macri}}}, \bibinfo {author} {\bibfnamefont {J.~E.}\ \bibnamefont
  {{Miraglia}}}, \ and\ \bibinfo {author} {\bibfnamefont {M.~S.}\ \bibnamefont
  {{Gravielle}}},\ }\href {\doibase 10.1364/JOSAB.20.001801} {\bibfield
  {journal} {\bibinfo  {journal} {Journal of the Optical Society of America B
  Optical Physics}\ }\textbf {\bibinfo {volume} {20}},\ \bibinfo {pages} {1801}
  (\bibinfo {year} {2003})}\BibitemShut {NoStop}%
\bibitem [{\citenamefont {Wolkow}(1935)}]{Volkov35}%
  \BibitemOpen
  \bibfield  {author} {\bibinfo {author} {\bibfnamefont {D.}~\bibnamefont
  {Wolkow}},\ }\href {\doibase 10.1007/BF01331022} {\bibfield  {journal}
  {\bibinfo  {journal} {Zeitschrift f\"ur Physik}\ }\textbf {\bibinfo {volume}
  {94}},\ \bibinfo {pages} {250} (\bibinfo {year} {1935})}\BibitemShut
  {NoStop}%
\bibitem [{\citenamefont {{Landau}}\ and\ \citenamefont
  {{Lifshitz}}(1965)}]{Landau65}%
  \BibitemOpen
  \bibfield  {author} {\bibinfo {author} {\bibfnamefont {L.~D.}\ \bibnamefont
  {{Landau}}}\ and\ \bibinfo {author} {\bibfnamefont {E.~M.}\ \bibnamefont
  {{Lifshitz}}},\ }\href@noop {} {\emph {\bibinfo {title} {Course of
  theoretical physics, Oxford: Pergamon Press, 1965}}}\ (\bibinfo {year}
  {1965})\BibitemShut {NoStop}%
\bibitem [{\citenamefont {{Perelomov}}\ \emph {et~al.}(1966)\citenamefont
  {{Perelomov}}, \citenamefont {{Popov}},\ and\ \citenamefont
  {{Terent'ev}}}]{PPT66}%
  \BibitemOpen
  \bibfield  {author} {\bibinfo {author} {\bibfnamefont {A.~M.}\ \bibnamefont
  {{Perelomov}}}, \bibinfo {author} {\bibfnamefont {V.~S.}\ \bibnamefont
  {{Popov}}}, \ and\ \bibinfo {author} {\bibfnamefont {M.~V.}\ \bibnamefont
  {{Terent'ev}}},\ }\href@noop {} {\bibfield  {journal} {\bibinfo  {journal}
  {Soviet Journal of Experimental and Theoretical Physics}\ }\textbf {\bibinfo
  {volume} {23}},\ \bibinfo {pages} {924} (\bibinfo {year} {1966})}\BibitemShut
  {NoStop}%
\bibitem [{\citenamefont {{Ammosov}}\ \emph {et~al.}(1986)\citenamefont
  {{Ammosov}}, \citenamefont {{Delone}},\ and\ \citenamefont
  {{Krainov}}}]{ADK86}%
  \BibitemOpen
  \bibfield  {author} {\bibinfo {author} {\bibfnamefont {M.~V.}\ \bibnamefont
  {{Ammosov}}}, \bibinfo {author} {\bibfnamefont {N.~B.}\ \bibnamefont
  {{Delone}}}, \ and\ \bibinfo {author} {\bibfnamefont {V.~P.}\ \bibnamefont
  {{Krainov}}},\ }in\ \href {\doibase 10.1117/12.938695} {\emph {\bibinfo
  {booktitle} {High intensity laser processes}}},\ \bibinfo {series} {Society
  of Photo-Optical Instrumentation Engineers (SPIE) Conference Series}, Vol.\
  \bibinfo {volume} {664},\ \bibinfo {editor} {edited by\ \bibinfo {editor}
  {\bibfnamefont {A.~J.}\ \bibnamefont {{Alcock}}}}\ (\bibinfo {year} {1986})\
  pp.\ \bibinfo {pages} {138--141}\BibitemShut {NoStop}%
\bibitem [{\citenamefont {{Goldstein}}\ \emph {et~al.}(2002)\citenamefont
  {{Goldstein}}, \citenamefont {{Poole}},\ and\ \citenamefont
  {{Safko}}}]{Goldstein02}%
  \BibitemOpen
  \bibfield  {author} {\bibinfo {author} {\bibfnamefont {H.}~\bibnamefont
  {{Goldstein}}}, \bibinfo {author} {\bibfnamefont {C.}~\bibnamefont
  {{Poole}}}, \ and\ \bibinfo {author} {\bibfnamefont {J.}~\bibnamefont
  {{Safko}}},\ }\href@noop {} {\emph {\bibinfo {title} {Classical mechanics
  (3rd ed.) by H.~Goldstein, C.~Poolo, and J.~Safko.~San Francisco:
  Addison-Wesley, 2002.}}}\ (\bibinfo {year} {2002})\BibitemShut {NoStop}%
\bibitem [{\citenamefont {Miller}(2007)}]{Miller71}%
  \BibitemOpen
  \bibfield  {author} {\bibinfo {author} {\bibfnamefont {W.~H.}\ \bibnamefont
  {Miller}},\ }\href {\doibase 10.1002/9780470143773.ch2} {\emph {\bibinfo
  {title} {Advances in Chemical Physics}}}\ (\bibinfo  {publisher} {John Wiley
  and Sons, Ltd},\ \bibinfo {year} {2007})\ pp.\ \bibinfo {pages} {69--177},\
  \Eprint
  {http://arxiv.org/abs/https://onlinelibrary.wiley.com/doi/pdf/10.1002/9780470143773.ch2}
  {https://onlinelibrary.wiley.com/doi/pdf/10.1002/9780470143773.ch2}
  \BibitemShut {NoStop}%
\bibitem [{\citenamefont {{Walser}}\ and\ \citenamefont
  {{Brabec}}(2003)}]{Walser03}%
  \BibitemOpen
  \bibfield  {author} {\bibinfo {author} {\bibfnamefont {M.~W.}\ \bibnamefont
  {{Walser}}}\ and\ \bibinfo {author} {\bibfnamefont {T.}~\bibnamefont
  {{Brabec}}},\ }\href {\doibase 10.1088/0953-4075/36/14/305} {\bibfield
  {journal} {\bibinfo  {journal} {Journal of Physics B Atomic Molecular
  Physics}\ }\textbf {\bibinfo {volume} {36}},\ \bibinfo {pages} {3025}
  (\bibinfo {year} {2003})}\BibitemShut {NoStop}%
\bibitem [{\citenamefont {{Spanner}}(2003)}]{Spanner03}%
  \BibitemOpen
  \bibfield  {author} {\bibinfo {author} {\bibfnamefont {M.}~\bibnamefont
  {{Spanner}}},\ }\href {\doibase 10.1103/PhysRevLett.90.233005} {\bibfield
  {journal} {\bibinfo  {journal} {Physical Review Letters}\ }\textbf {\bibinfo
  {volume} {90}},\ \bibinfo {eid} {233005} (\bibinfo {year}
  {2003})}\BibitemShut {NoStop}%
\bibitem [{\citenamefont {{Kay}}(2005)}]{Kay05}%
  \BibitemOpen
  \bibfield  {author} {\bibinfo {author} {\bibfnamefont {K.~G.}\ \bibnamefont
  {{Kay}}},\ }\href {\doibase 10.1146/annurev.physchem.56.092503.141257}
  {\bibfield  {journal} {\bibinfo  {journal} {Annual Review of Physical
  Chemistry}\ }\textbf {\bibinfo {volume} {56}},\ \bibinfo {pages} {255}
  (\bibinfo {year} {2005})}\BibitemShut {NoStop}%
\bibitem [{\citenamefont {Sch\"oller}\ \emph {et~al.}(1986)\citenamefont
  {Sch\"oller}, \citenamefont {Briggs},\ and\ \citenamefont
  {Dreizler}}]{Schoeller86}%
  \BibitemOpen
  \bibfield  {author} {\bibinfo {author} {\bibfnamefont {O.}~\bibnamefont
  {Sch\"oller}}, \bibinfo {author} {\bibfnamefont {J.~S.}\ \bibnamefont
  {Briggs}}, \ and\ \bibinfo {author} {\bibfnamefont {R.~M.}\ \bibnamefont
  {Dreizler}},\ }\href {http://stacks.iop.org/0022-3700/19/i=16/a=010}
  {\bibfield  {journal} {\bibinfo  {journal} {Journal of Physics B: Atomic and
  Molecular Physics}\ }\textbf {\bibinfo {volume} {19}},\ \bibinfo {pages}
  {2505} (\bibinfo {year} {1986})}\BibitemShut {NoStop}%
\bibitem [{\citenamefont {{Messiah}}(1973)}]{Messiah73}%
  \BibitemOpen
  \bibfield  {author} {\bibinfo {author} {\bibfnamefont {A.}~\bibnamefont
  {{Messiah}}},\ }\href@noop {} {\emph {\bibinfo {title} {{Quantum
  Mechanics}}}}\ (\bibinfo {year} {1973})\BibitemShut {NoStop}%
\bibitem [{\citenamefont {Dionissopoulou}\ \emph {et~al.}(1997)\citenamefont
  {Dionissopoulou}, \citenamefont {Mercouris}, \citenamefont {Lyras},\ and\
  \citenamefont {Nicolaides}}]{Dionissopoulou97}%
  \BibitemOpen
  \bibfield  {author} {\bibinfo {author} {\bibfnamefont {S.}~\bibnamefont
  {Dionissopoulou}}, \bibinfo {author} {\bibfnamefont {T.}~\bibnamefont
  {Mercouris}}, \bibinfo {author} {\bibfnamefont {A.}~\bibnamefont {Lyras}}, \
  and\ \bibinfo {author} {\bibfnamefont {C.~A.}\ \bibnamefont {Nicolaides}},\
  }\href {\doibase 10.1103/PhysRevA.55.4397} {\bibfield  {journal} {\bibinfo
  {journal} {Phys. Rev. A}\ }\textbf {\bibinfo {volume} {55}},\ \bibinfo
  {pages} {4397} (\bibinfo {year} {1997})}\BibitemShut {NoStop}%
\bibitem [{\citenamefont {{Fuller}}(1975)}]{Fuller75}%
  \BibitemOpen
  \bibfield  {author} {\bibinfo {author} {\bibfnamefont {R.~C.}\ \bibnamefont
  {{Fuller}}},\ }\href {\doibase 10.1103/PhysRevC.12.1561} {\bibfield
  {journal} {\bibinfo  {journal} {\prc}\ }\textbf {\bibinfo {volume} {12}},\
  \bibinfo {pages} {1561} (\bibinfo {year} {1975})}\BibitemShut {NoStop}%
\bibitem [{\citenamefont {{L{\'o}pez}}\ and\ \citenamefont
  {{Arb{\'o}}}(2019)}]{Lopez19}%
  \BibitemOpen
  \bibfield  {author} {\bibinfo {author} {\bibfnamefont {S.~D.}\ \bibnamefont
  {{L{\'o}pez}}}\ and\ \bibinfo {author} {\bibfnamefont {D.~G.}\ \bibnamefont
  {{Arb{\'o}}}},\ }\href {\doibase 10.1140/epjd/e2018-90528-5} {\bibfield
  {journal} {\bibinfo  {journal} {European Physical Journal D}\ }\textbf
  {\bibinfo {volume} {73}},\ \bibinfo {eid} {28} (\bibinfo {year}
  {2019})}\BibitemShut {NoStop}%
\bibitem [{\citenamefont {{Borb{\'e}ly}}\ \emph {et~al.}(2019)\citenamefont
  {{Borb{\'e}ly}}, \citenamefont {{T{\'o}th}}, \citenamefont {{Arb{\'o}}},
  \citenamefont {{T{\H{o}}k{\'e}si}},\ and\ \citenamefont
  {{Nagy}}}]{Borbely19}%
  \BibitemOpen
  \bibfield  {author} {\bibinfo {author} {\bibfnamefont {S.}~\bibnamefont
  {{Borb{\'e}ly}}}, \bibinfo {author} {\bibfnamefont {A.}~\bibnamefont
  {{T{\'o}th}}}, \bibinfo {author} {\bibfnamefont {D.~G.}\ \bibnamefont
  {{Arb{\'o}}}}, \bibinfo {author} {\bibfnamefont {K.}~\bibnamefont
  {{T{\H{o}}k{\'e}si}}}, \ and\ \bibinfo {author} {\bibfnamefont
  {L.}~\bibnamefont {{Nagy}}},\ }\href {\doibase 10.1103/PhysRevA.99.013413}
  {\bibfield  {journal} {\bibinfo  {journal} {\pra}\ }\textbf {\bibinfo
  {volume} {99}},\ \bibinfo {eid} {013413} (\bibinfo {year}
  {2019})}\BibitemShut {NoStop}%
\bibitem [{\citenamefont {{Dran}}\ and\ \citenamefont
  {{Arb{\'o}}}(2018)}]{Dran18}%
  \BibitemOpen
  \bibfield  {author} {\bibinfo {author} {\bibfnamefont {M.}~\bibnamefont
  {{Dran}}}\ and\ \bibinfo {author} {\bibfnamefont {D.~G.}\ \bibnamefont
  {{Arb{\'o}}}},\ }\href {\doibase 10.1103/PhysRevA.97.053406} {\bibfield
  {journal} {\bibinfo  {journal} {Phys. Rev. A}\ }\textbf {\bibinfo {volume}
  {97}},\ \bibinfo {eid} {053406} (\bibinfo {year} {2018})}\BibitemShut
  {NoStop}%
\end{thebibliography}%
%\nocite{*} 
% Produces the bibliography via BibTeX.
%---------------------------------------------------

\end{document}